\pdfoutput=1

\documentclass[11pt]{article}
\usepackage{booktabs}
\usepackage{graphicx}
\usepackage{amsmath}
\usepackage{multicol}
\usepackage{multirow}

\usepackage[final]{acl}

\usepackage{times}
\usepackage{latexsym}

\usepackage[T1]{fontenc}

\usepackage[utf8]{inputenc}

\usepackage{microtype}

\usepackage{inconsolata}

\usepackage{graphicx}

\def\ben{\textbf{Hindi-BEIR}}
\def\sys{\textbf{NLLB-E5}}

\usepackage{color}

\newcommand{\eat}[1]{} 

\eat{}

\title{\textbf{Benchmarking and Building Zero-Shot Hindi Retrieval Model with Hindi-BEIR and NLLB-E5}}

\author{
 \textbf{Arkadeep Acharya\textsuperscript{1}} \thanks{Work done as an intern at IBM Research},
 \textbf{Rudra Murthy\textsuperscript{2}},
 \textbf{Vishwajeet Kumar\textsuperscript{2}},
 \textbf{Jaydeep Sen\textsuperscript{2}},
\\
 \textsuperscript{1}Department of Computer Science and Engineering, Indian Institute of Technology Patna,\\
 \textsuperscript{2}IBM Research
\\
 arkadeep\_2101ai41@iitp.ac.in,\{rmurthyv,vishk024,jaydesen\}@in.ibm.com
}

\begin{document}
\maketitle
\begin{abstract}
Given the large number of Hindi speakers worldwide, there is a pressing need for robust and efficient information retrieval systems for Hindi. Despite ongoing research, comprehensive benchmarks for evaluating retrieval models in Hindi are lacking. To address this gap, we introduce the \textbf{Hindi-BEIR} benchmark,
comprising 15 datasets across seven distinct tasks. We evaluate state-of-the-art multilingual retrieval models on the Hindi-BEIR benchmark, identifying task and domain-specific challenges that impact Hindi retrieval performance. Building on the insights from these results, we introduce \textbf{NLLB-E5}, a novel multilingual retrieval model that leverages a zero-shot approach to support Hindi without the need for Hindi training data.  
We believe our contributions, which include the release of the \textbf{Hindi-BEIR} benchmark and the \textbf{NLLB-E5} model, will prove to be a valuable resource for researchers and promote advancements in multilingual retrieval models. The datasets from Hindi-BEIR are publicly available at \href{https://huggingface.co/collections/ArkaAcharya/datasets-667004c0dc348adcabc629be}{\textbf{Hindi-BEIR}}.The training and evaluation code for the NLLB-E5 model can be found on GitHub:  
\href{https://github.com/ArkadeepAcharya/NLLB-E5}{\textbf{NLLB-E5 Repository}}.
\end{abstract}

%

\section{Introduction}
\label{sec:intro}
Information retrieval (IR) models are indispensable in our digital age, enabling swift and accurate extraction of relevant data from vast amounts of information. These models enable quick and accurate retrieval of relevant data, thereby facilitating informed decision-making across various downstream applications and domains. In this modern era of LLMs, retrievers have become even more relevant and almost necessary in curbing hallucinations from Large Language Model (LLM) generations by means of much popular retrieval augmented generation (RAG)~\cite{lewis2021retrievalaugmented} methods. However, the focus of information retrieval (IR) research has predominantly centered on the English language, which can be attributed to the availability of comprehensive evaluation benchmarks, such as the BEIR~\cite{thakur2021beir}. Although, more recently, there have been efforts to build multi-lingual retrievers and associated evaluation benchmarks ~\cite{10.1162/tacl_a_00595,zhang-etal-2021-mr} on non-English languages, multi-lingual expansion of robust benchmarks is still a work in progress, with many languages like Hindi lacking a BEIR like a comprehensive benchmark. 

Hindi is one of the official languages of India\footnote{\url{https://en.wikipedia.org/wiki/Languages_of_India}}, a language of immense importance and global reach, with over half a billion speakers worldwide and ranking as the third most widely spoken language\footnote{\url{https://www.thecollector.com/what-are-the-most-spoken-languages-in-the-world/}}. However, from NLP research point of view, Hindi can be categorized as a low-resource language, lacking comprehensive resource and benchmarks to advance scientific research and this includes lack of a good retrieval benchmark too. Though there has been some efforts in recent time for the development of Indic language dataset for various NLP task like Question answering (\newcite{sabane2024breakinglanguagebarriersquestion}, \newcite{doddapaneni2023leavingindiclanguagebehind}) and Summarization (\newcite{datta-etal-2023-mildsum}, \newcite{ghosh2024medsumm}, \newcite{ghosh-etal-2024-healthalignsumm}), the evolution of Hindi or, in general, Indic language retrieval datasets has been rather ad-hoc, restricted to a small subset of domains, languages and tasks. Although a recent effort attempts to incorporate multiple Indic languages \cite{haq2023indicirsuite} for a retrieval benchmark, the absence of diversity in domains and tasks remains a major limitation, affecting the robust evaluation of the current state-of-the-art retriever models in practical applications, which often involve varying tasks and domains.

In this work, we posit that creating a BEIR-like diversified and robust benchmark is the necessary first step to track and advance any retrieval research effectively in a new language. Such a benchmark will be more useful than ad-hoc datasets with multiple languages. Thus, we choose the most widely spoken language among Indic languages, Hindi, as our pivot language to pioneer a large-scale diversified retrieval benchmark \textbf{Hindi-BEIR}, spanning 15 diverse datasets across \textbf{$7$} tasks. \textbf{Hindi-BEIR} aims for two key objectives:(1) Establish a standardized retrieval benchmark to assess, compare, and advance the state-of-the-art retriever models and (2) Provide workable insights into the potential research directions on building retrieval models for Hindi. With multiple tasks and domains in \textbf{Hindi-BEIR}, we posit that \textbf{Hindi-BEIR} will provide a more realistic and accurate assessment of retriever models for zero-shot usage in unseen domains and applications. 

On the choice of Hindi language it is important to note here, besides being a language of high usage, a retrieval benchmark in Hindi offers some unique challenges too such as:\\
    $\bullet$\textbf{ Script Difference:} Hindi uses the Devanagari script, which is fundamentally different from the Latin script used in English. This affects character encoding, text normalization, and processing. Existing tokenizers trained on English data may not handle Hindi text well, thus needing to evaluate various tokenization strategies too.
    
    

    $\bullet$\textbf{Grammatical Structure:} Hindi predominantly uses Subject-Object-Verb (SOV) word-order as opposed to the Subject-Verb-Object (SVO) word-order in English. Hindi words often include more inflections and agglutinations, affecting word tokenization and testing the robustness of retrieval models. 

    $\bullet$ \textbf{ Ambiguity:} In Hindi, some proper nouns can also function as common nouns. For instance, the name \textit{Lata}, a common female name, can also refer to \textit{creeper}, a common noun. A query like \textit{lata ko kaise saaph karen} (how to clean a creeper) can easily mislead lexicon-based retrieval systems and test a model's ability for word sense disambiguation.

The introduction of \textbf{Hindi-BEIR} benchmark lets us assess and compare the existing multi-lingual retriever models more accurately for their zero-shot performance in Hindi across diverse domains and tasks covered in the benchmark. Based on this performance comparison,   we notice that a key limitation of existing models is the need for a substantial amount of language-specific training data, which is often not available for low-resource languages such as Hindi. To address this problem, we propose \textbf{NLLB-E5}, a multilingual retrieval model, inspired by recent works like Langbridge \cite{yoon2024langbridgemultilingualreasoningmultilingual} and NLLB-LLM2Vec \cite{schmidt2024selfdistillationmodelstackingunlocks}, which distills information from a monolingual retrieval model into a multilingual retrieval model in a zero shot setup via a multi-lingual encoder. The key strength of \textbf{NLLB-E5} lies in its ability to perform distillation without reliance on multilingual training data, thereby eliminating the necessity for supervised multilingual datasets. 

We summarize our contributions as follows:
\begin{itemize}
    \item We introduce \textbf{Hindi-BEIR}, the first comprehensive retrieval benchmark in the Hindi language, encompassing 15 diverse datasets from 7 distinct tasks. This was accomplished through the harmonization of high-quality quality verified translated data from BEIR, the conversion of existing Hindi datasets for retrieval purposes, and the introduction of strategically generated synthetic data.

    \item We develop \textbf{NLLB-E5}, which combines a multilingual encoder with a monolingual retrieval model through knowledge distillation to develop a multilingual retrieval model without the need for any multilingual training data.

    \item We evaluate \textbf{NLLB-E5} on Hindi-BEIR and compare against existing multilingual models, which have been trained on multilingual data. We empirically demonstrate that \textbf{NLLB-E5} achieves strong performance on Hindi-BEIR, outperforming existing multilingual models without the need for multilingual training data.
\end{itemize}

\section{Related Works}

One of the known sources of testing retriever performance on Indic languages was to test on the Indic language subset contained in multi-lingual IR datasets such as MIRACL \cite{zhang2022making}, Mr.TyDi~\cite{zhang-etal-2021-mr} having instances from Indic languages such as Hindi, Bengali, and Telugu. 
However, two major drawbacks here were: (1) all of these subsets were specific to a domain, hence couldn't really be used as a robust benchmark, and (2) the Indic language-specific subsets of instances didn't contain enough data points to facilitate building neural retrieval models for Indic languages either. 

More recently, a relatively large-scale retrieval benchmark MS MARCO~\cite{DBLP:journals/corr/NguyenRSGTMD16} got translated into multiple languages to produce mMARCO~\cite{bonifacio2022mmarco}, including Hindi from the set of Indic languages. mMARCO was further extended to more Indic languages by IndicIRSuite~\cite{haq2023indicirsuite}, which is the most recent retrieval benchmark in Indic languages. However, the common problem for both mMARCO and  IndicMarco is that these benchmarks, too, are tied to a single type of retrieval task and web domain as was originally in MS MARCO and, therefore, do not provide a robust benchmark. 

In terms of Language specific benchmarks, \newcite{snegirev2024russian} introduce the ruMTEB Benchmark, an embedding model evaluation benchmark in Russian with Retrieval being a subtask, \newcite{valentini2024messirvelargescalespanishinformation} introduce the Spanish IR dataset while \newcite{wojtasik2024beirplzeroshotinformation} release the IR Benchmark in the Polish Language. In Indian languages such as Hindi, the Forum for Information Retrieval Evaluation (FIRE)\footnote{\url{https://fire.irsi.org.in/fire/2024/home}} has released numerous sub-tasks and datasets over the years to facilitate the benchmarking and development of information retrieval models. However, these datasets have predominantly been sourced from various news outlets and are not freely or openly accessible to the public. 

To the best of our knowledge, {\ben} presents the first comprehensive IR benchmark, which spans diverse domains and tasks and therefore, provides the first BEIR~\cite{thakur2021beir} equivalent comprehensive retrieval benchmark in Hindi.

For multilingual retrieval model development, the most promising works are from \newcite{chen2024bge} and \newcite{wang2024multilingual}, but both these models are heavily dependent on multilingual data and do not scale for languages and domains where data is not readily available in the target language.

\newcite{yoon2024langbridgemultilingualreasoningmultilingual} proposed an approach to equip existing LLMs with multilingual capability without any multilingual supervised data. They vertically stack a multilingual encoder and an LLM, adding a projection/alignment layer between them. The input first passes through the multilingual encoder, the representation from the multilingual encoder passes through the alignment layer and then the output from the alignment layer is passed as input to the LLM. They use English-labeled data to fine-tune the overall model and update only the embedding layers and the alignment layer. The model showed better multilingual capabilities.
\newcite{schmidt2024selfdistillationmodelstackingunlocks} extended the LLM2vec model \cite{behnamghader2024llmvec} by vertically stacking a NLLB model \cite{nllbteam2022languageleftbehindscaling} and LLM2vec model. Similar to LangBridge \cite{yoon2024langbridgemultilingualreasoningmultilingual}, they add a projection/alignment layer between the two and add LoRA parameters on the LLM. Different from LangBridge, they have a teacher model and a student model. They use LLM2vec as the teacher model and the NLLB-LLM2vec as the student model. The input is passed through both the teacher and the student models. They minimize the mean squared error between the final representations obtained from both the teacher and the student model. They train the model in two stages, the first stage being general alignment between the teacher and the student model, and, the second stage being task alignment. They show that the NLLB-LLM2vec model is able to obtain superior performance on various natural language understanding tasks.

Our {\sys} model is inspired by \newcite{yoon2024langbridgemultilingualreasoningmultilingual} who propose LangBridge, an approach which enables multilingual reasoning without multilingual supervised training and \newcite{schmidt2024selfdistillationmodelstackingunlocks} who opt for a distillation based approach and introduce NLLB-LLM2Vec capable of producing robust multilingual embeddings extending from NLLB encoder.

\section{\textbf{Hindi-BEIR} Retrieval Benchmark}

\begin{table}[!htb]
    \centering
    \resizebox{0.5\textwidth}{!}{%
    \begin{tabular}{lcrrrr}
    \toprule
    \multirow{2}{*}{\large\textbf{Dataset Name}} & 
    \multirow{2}{*}{\large\textbf{Tasks}} & 
    \multirow{2}{*}{\large\textbf{\#Corpus}} & 
    \multirow{2}{*}{\large\textbf{\# Queries}} & 
    \multicolumn{2}{c}{\textbf{Average \# words}}\\
    \cline{5-6}
    & & & & \textbf{Corpus} &\textbf{Query} \\
    \midrule

       ArguAna & Argument Retrieval & 7763 & 1194 & 159.20 & 178.20 \\
       FiQA-2018 & Question-Answering &48178 & 5924 & 118.02 & 15.23 \\
       TREC-COVID & Bio-Medical IR &76492 & 49 & 159.77 & 14.39 \\
       SCIDOCS & Citation-Prediction & 22050 & 850 & 169.88 & 12.82  \\
       SciFact & Fact-Checking & 2849 & 1099 & 164.37 & 19.44 \\
       Touch\'{e}-2020 & Argument Retrieval & 355273 & 49 & 351.47 & 8.58 \\
        NQ & Question Answering & 2595865  &2952 & 89.83 & 9.63  \\
        FEVER & Fact Checking & 5362876 & 120075 & 88.86 & 9.48\\
        Climate-FEVER & Fact Checking & 5362911 & 1499 & 88.86 & 24.40 \\
        \midrule

        CC News Retrieval & News Article Retrieval& 5005483 & 49699 & 272.20 & 9.30 \\
        Sangraha-IR & Question Answering & 350000 & 9744 & 308.13 & 30.91 \\
        \midrule

        MIRACL & Passage Retrieval & 506264 & 350 & 66.49 & 10.42 \\
        IndicQARetrieval &Question Answering & 261 & 1544 & 480.66 & 10.88\\
        mMARCO &Passage Retrieval& 8841823 & 6980 & 64.40 & 6.46 \\
       WikiPediaRetreival & Question Answering & 13500 & 1500 & 77.19 & 9.54\\
    \bottomrule

    \end{tabular}%
    }
    \caption{Statistics of the Dataset in the Hindi-BEIR Benchmark showing the tasks which each dataset covers and the number of corpus and  query in the evaluation set of each dataset in the Hindi-BEIR Benchmark.}
    \label{tab:dataset_stats}
\end{table}

Table \ref{tab:dataset_stats} summarises the various datasets included in the {\ben} benchmark, along with the number of documents, queries, and domain information.

To ensure that the {\ben} Benchmark is comprehensive, challenging, and accessible to the public for future research and evaluation, we adhered to the following objectives:\textbf{1) Diverse Domains and Tasks:} We include datasets from diverse domains like Wikipedia and news articles to niche domains like scientific publications, finance to test the robustness and generalization ability of the retrieval models. Additionally, a good retrieval model should handle documents and queries of varying lengths equally well. We have ensured that the datasets included in the Hindi-BEIR benchmark exhibit a wide range of query and document lengths among the datasets. \textbf{2) Difficulty Level:} We ensure that the datasets are challenging and systems relying solely on lexical overlap have a hard time retrieving the correct document. \textbf{3) Public Availability: } All datasets curated in {\ben} Benchmark have user-friendly licenses and are publicly available at \href{https://huggingface.co/collections/ArkaAcharya/datasets-667004c0dc348adcabc629be}{Hindi-BEIR}

We discuss the {\ben} benchmark creation in the following subsections.
Please refer to Appendix \ref{subsec:dataset_desc} for a more detailed analysis of each dataset.
\subsection{Translating English Datasets to Hindi:}
\label{subsec:translation}
We translate a subset of the existing English datasets from the BEIR benchmark \cite{thakur2021beir} into Hindi. This approach was necessitated by the need to ensure comprehensive coverage across multiple domains and to maintain a high level of data quality and complexity. We utilized the Indic-Trans2 model \footnote{Our choice for IndicTrans2 over other translation models has been discussed in \ref{subsec:faq} } \cite{gala2023indictrans}, a multilingual NMT model supporting translations across all 22 scheduled Indic languages (including English). We employ the back-translation technique to retain good translations. Specifically, given an English query/document we translate it to Hindi. This Hindi-translated query/document is then translated back to English. We calculate the Chrf(++) score \cite{popovic-2017-chrf} between the original English query/document and the backtranslated English query/document. We retained only those translations with a Chrf(++) score exceeding a threshold. We empirically set the threshold to 50 after manually verifying the translation quality of texts obtained from different thresholds. 

All entries that fell below this threshold were removed. If an entry was part of the corpus and had an associated query, the query was also discarded. Likewise, any corresponding records in the relevancy mapping file were also deleted.

This strategy enables us to leverage the wealth of existing high-quality datasets in English while making them accessible and useful for Hindi language information retrieval tasks. Nine out of the fifteen datasets, which include Arguana, FiQA-2018, TREC-COVID, SCIDOCS, SciFact, Touch\'e-2020, NQ, FEVER, and Climate-FEVER were created by this method.

\begin{figure*}
    \centering
    \includegraphics[width=1\linewidth]{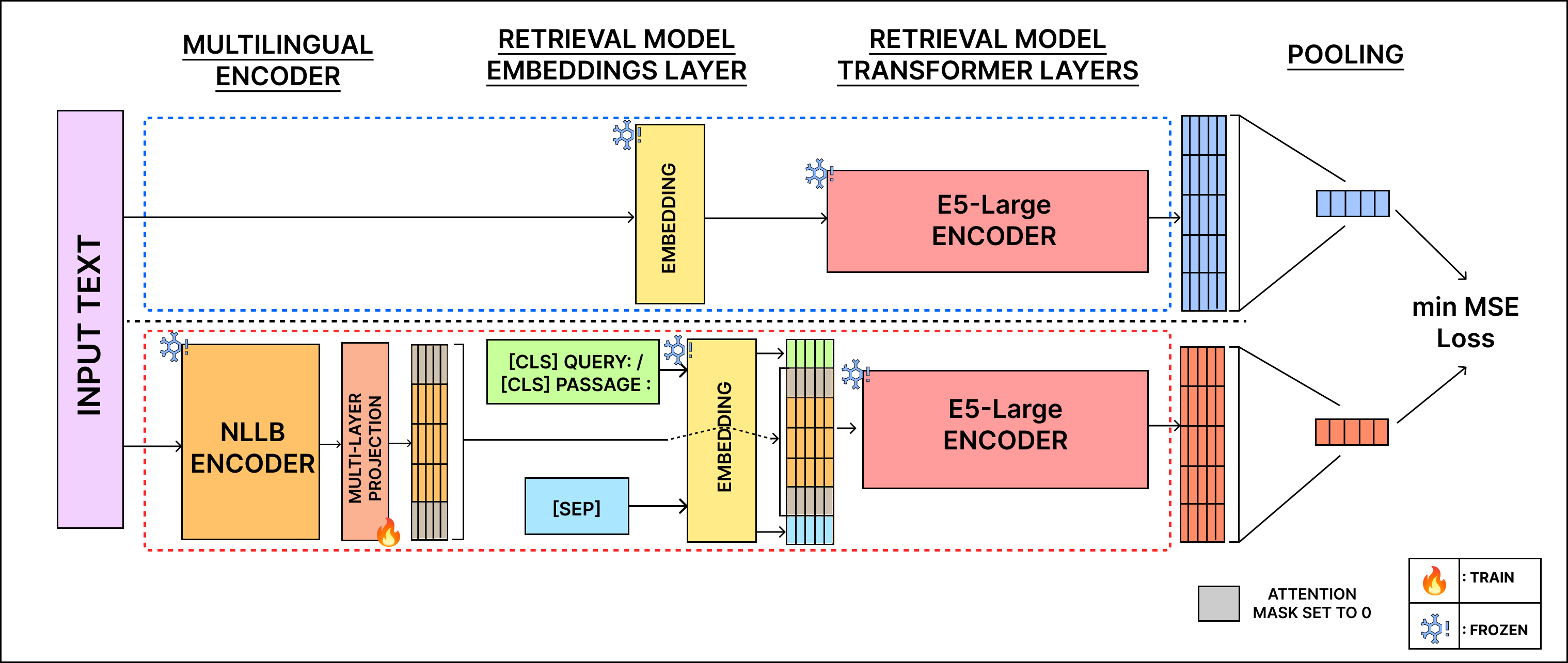}
    \caption{A comprehensive overview of our training methodology. The portion above the dotted lines represents the teacher model (enclosed by a blue dotted box), while the portion below represents our proposed NLLB-E5 student model (enclosed by a red dotted box). We use NLLB as a multilingual encoder and train the Linear projection layer and the Lora adapters by forcing its outputs to match that of the teacher model via the Mean Squared Error (MSE).}
    \label{fig:architecture}
\end{figure*}

\subsection{Using existing datasets for Retrieval task}
\label{cc_news}
We create Hindi CC News Retrieval and Sangraha-IR datasets from existing datasets. We detail the process we followed for each of the datasets as follows:

\textbf{Hindi CC News Retrieval :} This is a cross-lingual dataset derived from the Hindi subset of the Multilingual CC News dataset\footnote{\url{https://huggingface.co/datasets/intfloat/multilingual_cc_news}}, which contains 7,444,584 data points comprising Hindi news articles and their corresponding titles. The article text, which is in Hindi, serves as the corpus to be retrieved. At the same time, the English news titles act as queries, representing a practical and common real-world scenario that needs the models to have language-agnostic knowledge across both Hindi and English.



We selected only those title-article pairs where the title never appears in the article text, thus reducing the dataset size to 451,803. This ensures no lexical overlap between the title and the article, making the retrieval task challenging as the model needs to rely on cross-lingual cues. From these filtered data points, we randomly selected 49,699 queries to form the test split, which has been incorporated into the Hindi-BEIR benchmark. Additionally, we release training and validation splits consisting of 301,578 and 100,526 query-corpus pairs, respectively. The train, test, and validation splits collectively encompass all data points obtained after the second filtering step.

\textbf{Sangraha-IR :}This dataset builds on top of the Sangraha Dataset \cite{khan-etal-2024-indicllmsuite}. We handpicked 350,000 entries from the Sangraha-verified subset, focusing on topics that truly represent Indian culture, such as agriculture, Bollywood, cricket, and other similar areas that capture the heart of India. We specifically chose the Sangraha-verified corpus as it consists of scraped data from "human-verified" Websites, OCR-extracted data from high-quality Indic language PDFs, transcribed data from various Indic language videos, podcasts, movies, courses, etc, and truly captures the essence of the language. To construct a set of relevant queries, we randomly selected 10,000 data points and used the Google Gemini-1.5-flash model to generate corresponding queries \footnote{Please refer to the Appendix for the prompt used} for these documents. Queries that did not adhere to the provided instructions or deviated from the desired format were excluded, resulting in a final dataset of 9,744 queries and 350K candidate documents.

\subsection{Subset from existing Multilingual Datasets}
We also include 4 publicly available datasets, namely MIRACL (dev split)\cite{zhang-etal-2021-mr},
mMarco \cite{bonifacio2022mmarco}, IndicQARetrieval (developed by modifying the IndicQA Dataset \cite{doddapaneni2023leaving}) and WikiPediaRetrieval \footnote{\url{https://huggingface.co/datasets/ellamind/wikipedia-2023-11-retrieval-multilingual-queries}}.

For translation, we specifically selected a subset of BEIR that focuses on factual and general knowledge content, avoiding topics with strong cultural or language-specific nuances that might be lost in translation. However, to ensure the inclusion of real-world data that captures the linguistic intricacies of Hindi, we carefully incorporate datasets such as Sangraha-IR, CC News Retrieval, MIRACL, IndicQA Retrieval, and Wikipedia Retrieval. These datasets, sourced from manually curated or web-scraped content, effectively preserve the cultural and language-specific nuances essential for robust evaluation.

\section{NLLB-E5: multilingual retriever}
\label{methodology}
In this section, we build a strong multilingual retrieval model based on the approach proposed by \newcite{yoon2024langbridgemultilingualreasoningmultilingual} and \newcite{schmidt2024selfdistillationmodelstackingunlocks}. 



Figure~\ref{fig:architecture} shows the architecture of \textbf{NLLB-E5} with the different components and data flow between them. We build \textbf{NLLB-E5} on the back of two components (1) A multilingual encoder NLLB \cite{nllbteam2022languageleftbehindscaling} and (2) a monolingual retriever E5 \cite{wang2024textembeddingsweaklysupervisedcontrastive}. We bank on the multilingual encoder to project semantically similar sentences across languages to a shared representation space. We hypothesize that embeddings coming from such a language-agnostic shared representation space can act as a strong prior that can be easily aligned for producing multilingual task-specific embeddings. Moreover, given the multilingual nature of the embedding coming directly from the multilingual encoder, the aforementioned task-specific alignment is language agnostic and, therefore, can be done with task-specific data in a high-resource language, too. 

Figure \ref{fig:architecture} demonstrates how we follow the above intuition in designing \textbf{NLLB-E5}.  We use NLLB~\newcite{nllbteam2022languageleftbehindscaling} as the strong multilingual encoder that produces robust multilingual embeddings owing to its translation capabilities for over 200 languages. To align these embeddings for the retrieval task, we learn a projection layer from the multilingual NLLB encoder to a monolingual retriever E5. The alignment happens by learning the projection layer parameters between NLLB and E5 through distillation from E5 embeddings, which are specifically tuned for retrieval tasks. Since this alignment to the retrieval task is agnostic to a specific language, we use English data and monolingual retriever E5 to do the distillation, eliminating the need for Hindi training data. We chose E5-large as the teacher and student model coupled with NLLB owing to its satisfactory retrieval performance on the BEIR benchmark.

\section{Training Methodology}
The {\sys} model is constructed by integrating the monolingual E5 model \cite{wang2024textembeddingsweaklysupervisedcontrastive} atop the multilingual encoder from NLLB \cite{nllbteam2022languageleftbehindscaling}, connected via a learnable projection layer. The parameters of both the monolingual and multilingual base models remain frozen throughout training. To enable adaptation, we introduce $W \in {R}^{n \times d}$, a linear projection layer mapping the $n$-dimensional output space of the multilingual encoder to the $d$-dimensional input space expected by the E5 model.

We employ knowledge distillation \cite{hinton2015distillingknowledgeneuralnetwork} to transfer the capabilities of the monolingual E5 model to the {\sys} model, facilitating multilingual sentence representation learning. Specifically, the original E5 model serves as the \emph{teacher}, producing high-quality English embeddings, while the {\sys} model, which includes the learnable Linear projection $W$, acts as the \emph{student}. By minimizing the discrepancy between the teacher and student embeddings, the student model learns to approximate the E5 model's performance across Hindi and other languages supported by NLLB. The frozen multilingual encoder ensures effective cross-lingual transfer, enabling generalization across all NLLB supported languages. \footnote{We also experiment with LoRA adapters added on top of the E5 model. Details have been discussed in the Appendix \ref{lora}}

\subsection{Training Process}

The training procedure consists of the following steps:

\textbf{Get the Teacher Model Embeddings:}
Given an input English sentence \( S \), we first tokenized it using the E5 tokenizer, yielding a sequence of tokens $\{t_1, t_2, \dots, t_n\}$.
These tokens are passed through the teacher E5 model, with frozen parameters, denoted as \( \mathcal{M}_T \). The model produces a sequence of token-level representations for the tokens:
\[
\{X_1, X_2, \dots, X_n\} = \mathcal{M}_T(\{t_1, t_2, \dots, t_n\})
\]
where \( X_i \in {R}^d \) represents the \( d \)-dimensional embedding of the token \( t_i \) produced by the model \( \mathcal{M}_T \).
To obtain the final sentence-level embedding \( \vec{X}_T \), we apply average pooling over the token-level representations, which act as the final embedding for the sentence \( S \) obtained from the \emph{teacher model}.

\textbf{Get Student Model Embeddings:}
The input sentence is tokenized using the NLLB tokenizer, producing a sequence of tokens $\{t_1, t_2, \dots, t_m\}$. The frozen multilingual encoder processes this token sequence, generating token-level representations in the multilingual semantic space $\{H_1, H_2, \dots, H_m\}$. To align with the input structure expected by the E5 model, the token sequence is modified by prepending a token embedding corresponding to either \texttt{[CLS] query:} (for queries) or \texttt{[CLS] passage:} (for corpus), obtained from the E5 embedding layer (referred from here on as $H_{\emph{prefix}}$), along with the representation of \texttt{[SEP]}(referred from here on as $H_{\emph{postfix}}$) which is appended at the end. This is  projected through a learnable layer \( W \) as $
\{H'_1, H'_2, \dots, H'_m\} = W \{H_{\emph{prefix}},H_1, H_2, \dots, H_m, H_{\emph{postfix}}\}$
This projected sequence is then passed through the student E5 model, which outputs the token-level representations $\{X'_1, X'_2, \dots, X'_m\}$
Finally, average pooling is applied over these token-level representations to obtain the final sentence-level embedding \( \vec{X}_S \) from the student model.

\textbf{Minimizing the Loss Function.} To align the student model with the teacher, we minimize the Mean Squared Error (MSE) between the teacher and student embeddings, $\vec{X}_T$ and $\vec{X}_S$, respectively, for a batch of sentences. 


This optimization encourages the student to generate multilingual embeddings that are aligned with the high-quality English embeddings produced by the teacher. 

\eat{
\subsection{Training Data}
We aim to align our student model with the teacher model for multiple retrieval tasks by exposing it to diverse datasets. To accomplish this, we trained our model on a subset of the datasets used for training sentence transformer models. We randomly selected 100,000 samples from each dataset to create the final training set. The datasets chosen are i) Arxiv ii) Natural Questions (NQ) iii) HotpotQA iv) Stack Exchange - Duplicate questions (titles) v) Stack Exchange - (Title, Body) pairs vi) Stack Exchange - (Title+Body, Answer) pairs vii) Stack Exchange - (Title, Answer) pairs viii) Stack Exchange - Duplicate questions (bodies) ix) Stack Exchange duplicate questions (titles+bodies) x) SQuAD 2.0 xi) S2ORC Citation - pairs (Abstracts) xi) S2ORC (Title, Abstract) xii) SearchQA xiii) PubMed (Title, Abstract) xiv) SPECTER xv) FEVER xvi)
Wikipedia Sections xvii) Stack Exchange Math xviii) Stack Overflow Posts xix) Wikipedia 
}
\section{Experiments}
\label{sec:expr}
This section provides a detailed overview of our experiments on \textbf{Hindi-BIER} datasets, including our experimental setup and the baseline models used in our experimentation.

\subsection{Benchmark and Evaluation Metric}
We use \textbf{Hindi-BEIR} as the target benchmark for all our experiments. We evaluate the models across all $15$ datasets in \textbf{Hindi-BEIR} and compare their performance at the dataset level too for finer insights. 

Following BEIR~\cite{thakur2021beir} benchmark standard, we use Normalised Discounted Cumulative Gain (NDCG)~\cite{Jrvelin2002CumulatedGE}, more specifically NDCG@10 as our evaluation metric. NDCG is known to be a more robust metric than simple recall in the IR community because it is purposefully designed to be sensitive towards ranking of the retrieved results.

\subsection{Implementation Details of NLLB-E5}
To make \textbf{NLLB-E5} produce robust, generalizable embeddings across different kind of retrieval tasks and domains, we rely on using diverse data to learn NLLB and E5 alignment during training. We followed a similar approach as taken by generic text-embedding models and used a subset of data from the huge pool of ~1B diverse data curated for training sentence transformer models~\cite{sentence-transformers}. We randomly selected 100,000 samples from each dataset to create the final training set. The datasets chosen are i) Arxiv ii) Natural Questions (NQ) iii) HotpotQA iv) Stack Exchange - Duplicate questions (titles) v) Stack Exchange - (Title, Body) pairs vi) Stack Exchange - (Title+Body, Answer) pairs vii) Stack Exchange - (Title, Answer) pairs viii) Stack Exchange - Duplicate questions (bodies) ix) Stack Exchange duplicate questions (titles+bodies) x) SQuAD 2.0 xi) S2ORC Citation - pairs (Abstracts) xi) S2ORC (Title, Abstract) xii) SearchQA xiii) PubMed (Title, Abstract) xiv) SPECTER xv) FEVER xvi)
Wikipedia Sections xvii) Stack Exchange Math xviii) Stack Overflow Posts xix) Wikipedia 

We train the {\sys} model for 10 epochs, with a learning rate of 2e-4 and a linear scheduler for learning rate adjustment with FP16 precision. We chose the best checkpoint based on the evaluation result on the MsMARCO dev set. We ran all our experiments using Nvidia A100 80GB GPUs. Evaluating the {\sys} model using 4 A100 80GB GPUs on the {\ben} benchmark took 24 hours.

\begin{table*}[!htb]
    \centering
    \resizebox{0.9\textwidth}{!}{%
    \begin{tabular}{lrrrrrrrr}
    \toprule
      \textbf{Dataset Name} & \multicolumn{1}{c}{\begin{tabular}[c]{@{}c@{}}\textbf{BGE-M3}\\ (567M)\end{tabular}}&  \multicolumn{1}{c}{\begin{tabular}[c]{@{}c@{}}\textbf{mE5-Base}\\ (110M)\end{tabular}} &  \multicolumn{1}{c}{\begin{tabular}[c]{@{}c@{}}\textbf{mE5-Large}\\ (335M)\end{tabular}} & \textbf{LASER} &  \multicolumn{1}{c}{\begin{tabular}[c]{@{}c@{}}\textbf{LaBSE}\\ (471M)\end{tabular}}&  \textbf{BM-25} & 
      \textbf{NLLB-E5 (3.3B)}\\
    \midrule
       ArguAna & 53.81 &49.96 & \underline{54.77} & 11.27 & 32.90&  43.75 & \textbf{57.68} \\
       FiQA-2018 & 25.89 & 22.38 &\underline{27.33}& 1.58 & 7.23&  16.57 & \textbf{34.41}\\
       TREC-COVID & 64.60 &62.42 &\textbf{70.80}& 3.95 & 29.94&  52.30 & \underline{70.12}  \\
       SCIDOCS & \underline{14.24} & 10.42 &11.32& 0.59 & 6.95 &  11.40 & \textbf{17.82} \\
       SciFact & 52.39& 51.50 &55.92& 5.37 & 33.42  &  \underline{60.80} & \textbf{64.34} & \\
       Touch\'{e}-2020 & 26.68 & \underline{27.44}&26.89 & 1.06 & 6.82  & \textbf{33.59} & 25.35\\
        NQ & 39.15 & \underline{44.10}&\underline{44.10}& 0.49 & 9.36 &  16.79 & \textbf{53.04} \\
        FEVER & \underline{66.91} & 32.87&39.36 & 0.19 & 8.27 &  40.57& \textbf{71.82}\\
        Climate-FEVER & \underline{23.71} & 5.93& 8.22& 0.28 & 3.72 & 14.00& \textbf{25.66}\\
        \midrule
        CC News Retrieval & \underline{34.40} & 20.81& \textbf{35.00}& 0.52 & 5.63 & 0.01& 31.73 \\
        Sangraha-IR &\underline{44.85} &41.12 &\textbf{47.91}&3.76 &21.76 &30.92 & 42.36 \\
        \midrule
        MIRACL & \textbf{59.34} & 58.11 & \underline{59.24}& 0.69 & 13.76 &  40.98 & 53.56\\
        IndicQARetrieval & \textbf{69.92} & \underline{67.11}& \underline{67.11} & 21.28 & 46.85 &  74.02 & 61.94\\
        mMARCO & 29.49& 29.94 &\underline{30.92}& 0.48 & 6.98 &  16.53 & \textbf{33.03}\\
       WikiPediaRetrieval & \textbf{87.38} & 84.40 &\underline{86.71}& 0.03 &61.28 &  82.54 & 85.65 \\
       \midrule
       \textbf{Average} & \underline{46.18} & 40.57 &44.37& 3.44&19.66& 35.65 & \textbf{48.57}\\
    \bottomrule    
    \end{tabular}%
    }
    \caption{NDCG@10 scores of the existing multilingual model and best-performing NLLB-E5 on Hindi-BEIR datasets. The best-performing model has been highlighted as \textbf{bold} while the second-best model has been \underline{underlined}.}
    \label{tab:results}
\end{table*}

\subsection{Baseline Models}
We evaluate our Hindi-BEIR dataset on the following models:

\begin{enumerate}
\item The \textbf{NLLB-E5 variations:} We develop and evaluate NLLB-E5  using 600M, 1.3B and 3.3B parameter versions of the NLLB model and compare their performance\footnote{\url{https://huggingface.co/facebook/nllb-200-distilled-1.3B}, \url{https://huggingface.co/facebook/nllb-200-distilled-600M}, \url{https://huggingface.co/facebook/nllb-200-3.3B}}. 

\item The \textbf{BGE-M3} \cite{chen2024bge} model was pre-trained on a large multilingual and cross-lingual unsupervised data and subsequently fine-tuned on high-quality multilingual retrieval datasets using a custom loss function based on the InfoNCE loss function. BGE-M3 supports a context length of 8,192. 

\item\textbf{Multilingual E5 (mE5)} \cite{wang2024multilingual} was developed by continually pre-training the E5 model \cite{wang2024text} on a large multilingual corpus using a weakly supervised contrastive pretraining method with InfoNCE contrastive loss \newcite{oord2019representation}. It was then fine-tuned on high-quality labelled multilingual datasets for retrieval tasks. mE5 has a context length of 512 tokens.

\item \textbf{LASER} \cite{Artetxe_2019} focuses on universal language agnostic sentence embeddings across 93 languages. LASER uses a language-agnostic BiLSTM encoder architecture trained on parallel corpora from different languages without any retrieval-specific fine-tuning. 

\item \textbf{LaBSE} \cite{feng2022languageagnostic} uses a dual encoder model using BERT to obtain language-agnostic sentence embedding without any retrieval-specific fine-tuning. It has a maximum context length of 256 tokens.
\item \textbf{BM-25} \cite{Amati2009} is a ranking function in information retrieval based on estimating relevance by combining term frequency and document length normalization.
\end{enumerate}
We extend the MTEB benchmark code repository\footnote{\url{https://github.com/embeddings-benchmark/mteb}} to include Hindi-BEIR and evaluate the models discussed above.

\section{Results and Analysis}
In this section, we compare the performance of the aforementioned baselines with our proposed {\sys} model on \textbf{Hindi-BEIR} benchmark and highlight our key findings from the results.\footnote{We also evaluated NLLB Encoder trained using distillation approach as used for NLLB-E5. Details has been discussed in \ref{nllb_baseline}}

\begin{table}[!htb]
    \centering
    \tiny
    \resizebox{0.5\textwidth}{!}{%
    \begin{tabular}{lcccc}
    \toprule
         \textbf{Dataset Name}& 
         \multicolumn{1}{c}{\textbf{\begin{tabular}[c]{@{}c@{}}600M\\ (distilled)\end{tabular}}}
         & \multicolumn{1}{c}{\textbf{\begin{tabular}[c]{@{}c@{}}1.3B\\ (distilled)\end{tabular}}}& \textbf{1.3B}& \textbf{3.3B} \\
         \midrule
         Arguana & 55.23 & 57.24& 56.50 &57.68\\
         FiQa-2018 & 32.83 & 34.19 & 33.92 & 34.41 \\
         TREC-COVID & 70.07 & 70.53 & 70.54 & 70.12\\
         SCIDOCS & 17.53 & 18.23 & 17.67 & 17.82\\
         SciFact & 63.93 & 64.36& 64.35 & 64.34 \\
         Touch\'{e}-2020 & 25.89 & 25.21& 26.13 & 25.35\\
         NQ & 50.14 & 53.38& 52.77 & 53.04 \\
         FEVER & 66.79 & 71.04& 69.02 & 71.82\\
         Climate-FEVER & 25.12 & 22.51& 23.79 & 25.66\\
         \midrule
         CC News Retrieval & 29.11 & 31.91& 30.93 & 31.73 \\
         Sangraha-IR & 40.51 & 43.30& 42.84 & 42.36\\
         \midrule
         MIRACL & 50.38 & 52.96& 52.88 & 53.56\\
         IndicQARetrieval & 60.47 & 62.10& 61.69 & 61.94\\
         mMARCO & 31.63 &34.03 & 33.15 & 33.03\\
         WikiPediaRetrieval & 84.89 & 86.20& 86.22 & 85.65\\
         \midrule
         \textbf{Average} & 46.97 & 48.48& 48.16 &48.57\\
         \bottomrule

    \end{tabular}
    }
    \caption{NDCG@10 scores of NLLB-E5 models using NLLB encoders of different parameter sizes on \textbf{Hindi-BEIR} benchmark}
    \label{tab:parameter_ablation}
\end{table}

\eat{
\textbf{{\sys} outperforms multilingual retrievers on {\ben}:} Table \ref{tab:results} showcases the strong performance of the NLLB$_{1.3B}$-E5 model on the Hindi-BEIR Benchmark. Specifically, the NLLB$_{1.3B}$-E5 model attains the highest scores on the Hindi versions of ArguAna, FiQA-2018, NQ, SCIDOCS and SciFact, FEVER, Climate-FEVER and mMARCO and performs 2nd best on Hindi TREC-COVID dataset. In comparison to its direct competitor, mE5-Large, the NLLB-E5 model demonstrates a substantial edge in fact-checking datasets such as FEVER, Climate-FEVER and SciFact, resulting in a significant boost of 82.47 \%, 212 \% and 15.06 \% on the respective datasets. Moreover, the NLLB-E5 model's performance metrics closely mirror the best-performing model on CC News Retrieval and WikiPediaRetrieval. }

\textbf{{\sys} outperforms multilingual retrievers on {\ben}:} Table \ref{tab:results} shows NLLB$_{1.3B}$-E5 model achieves better average compared to all other baselines on Hindi-BEIR. This is significant because we had state-of-the-art multilingual retriever models such as BGE-M3 and mE5 as our baselines in addition to LASER and LaBSE. It is important to note here that all these multilingual embedding models have seen a huge amount of multilingual data in their training, including Hindi. In contrast, {\sys} is tuned only on English language data for learning retrieval alignment multilingual embeddings from NLLB encoder. 

As we can see in Table \ref{tab:results} {\sys} performs the best on almost all of BEIR subsets of Hindi-BEIR. More specifically, the {\sys} model attains the highest scores on the Hindi versions of ArguAna, FiQA-2018, NQ, SCIDOCS and SciFact, FEVER, Climate-FEVER and mMARCO and performs 2nd best on Hindi TREC-COVID dataset. In comparison to its direct competitor, mE5-Large, the NLLB-E5 model demonstrates a substantial edge in fact-checking datasets such as FEVER, Climate-FEVER and SciFact, resulting in a significant boost of 82.47 \%, 212 \% and 15.06 \% on the respective datasets.  For non-BIER subsets too, NLLB$_{1.3B}$-E5 easily outperforms LASER and LaBSE and performs competitively with mE5 and BGE-M3.

\textbf{{\sys} outperforms BM25 on {\ben}:} BM25 is known to be a strong baseline on BEIR~\cite{thakur2021beir} benchmark. In fact, as seen in Table \ref{tab:results}, BM25 indeed acts as a better baseline than LASER and LaBSE too. However, {\sys} outperforms BM25 on 13/15 datasets in the Hindi-BEIR benchmark, achieving a gain of $~+37\%$ on average. This gain justifies that we can treat {\sys} as a robust neural model much better than lexical retrievers such as BM25.

BM25 has very poor performance on CC News Retrieval datasets only. The CC News Retrieval dataset was intentionally designed to test the ability of multilingual dense retrieval models to learn language-agnostic embeddings for retrieval. This dataset presents a scenario where the query is in English, and the corpus is in Hindi, which needs to move beyond token-level matching. This explains the poor performance of BM25 due to the lack of overlap between English and Hindi tokens.

\textbf{Variation in Performance Across Different Tasks and Domains:} One of the important aspects of building Hindi-BEIR was to see how a retriever adapts to different domains and tasks. As we can see, mE5 specifically struggles in fact-checking datasets, while BGE-M3 and NLLB-E5, although not great, still fare better for fact-checking tasks. All the models perform quite poorly for the citation prediction tasks on SCIDOCS. Also, for niche domains such as Finance and Climate, we see a sharp drop for all the models, prompting the need for focused research in these areas.

 \textbf{Effect of Encoder Param Size: } Table \ref{tab:parameter_ablation} presents the NDCG@10 scores for the NLLB model's encoder across different parameter sizes. As the table shows, performance generally improves as the parameter size increases, though minor anomalies exist across specific datasets. Notably, the performance gain plateaus beyond the 1.3B parameter model, with only a marginal improvement of 0.09\% when comparing the NLLB-E5 model's 1.3B (distilled) encoder to that of the 3.3B parameter version. This possibly indicates that {\sys} is not lacking in model capacity yet and thus can possibly be improved further by showing more volumes of data from diverse tasks and domains.

\section{Conclusion and Future Work}
In this work, we introduced {\ben}, a comprehensive benchmark for Hindi language information retrieval. The benchmark consists of 15 datasets, with a corpus size exceeding 27 million and a query size of over 200K, covering a diverse range of 7 tasks. This provides the first-of-its-kind robust benchmark for assessing the performance of retriever models in Hindi.  Our experimental results offer insights into the strengths and limitations of current multilingual retrievers on {\ben}, indicating the pressing need for further research in this domain.  Additionally, we also proposed the {\sys} model, a novel architecture for multilingual retriever tuning without needing multilingual training data, capable of handling data from over 200 languages, including Hindi, which outperformed existing multilingual models on Hindi-BEIR.

Future work will focus on expanding the Hindi-BEIR benchmark to include more diversity by curating additional domains such as Law and Medicine. We also plan to extend this benchmark to cover languages beyond Hindi and explore alternative multilingual encoders to optimize representation. We believe our work will have a lasting impact in developing inclusive and scalable information retrieval systems across diverse languages.

\section{Limitations}
\textbf{The queries are AI generated etc.}
While the {\ben} Benchmark and {\sys} model provides valuable advancements, we acknowledge several limitations. One limitation of the {\ben} Benchmark is that it may not fully capture the breadth of tasks where retrieval models are essential. The current scope, while extensive, lacks coverage in critical domains such as Law and Medicine, which are key areas we plan to include in future expansions. Additionally, although designed primarily for Hindi, the benchmark’s extension to other languages is an area that remains unexplored in this work.

The {\sys} model, while showing promise, exhibits sub-optimal performance on certain English tasks. This limitation indicates a need to explore alternative multilingual encoders to improve language performance. Additionally, we realise that the current model would architecture struggle with long texts due to a context length limitation of 508 tokens, which poses challenges in tasks requiring extended context handling. Further, we would also like to extend and validate this idea for other low-resource languages in the future.

\bibliography{custom}

\appendix

\section{Appendix}
\label{sec:appendix}
\subsection{Dataset License}
\label{license}
The datasets will be publicly released with the same license as the parent dataset and will be made available for research purposes.
\subsection{Dataset Description}
\label{subsec:dataset_desc}
\subsubsection{ArguAna}
\begin{enumerate}
\item \textbf{Task definition:} Derived from the work by \newcite{wachsmuth-etal-2018-retrieval} , the task is to retrieve the best counterargument, given an argument. Translations of arguments and counter-arguments from online debates constitute the corpus, while the translations of arguments in the original test split, after going through the filtration process based on Chrf++ scores (refer to \ref{subsec:translation}), constitute the queries.

    \item \textbf{Domain :} Misc.
\end{enumerate}

\begin{figure}[hbt!]
    \centering
    \includegraphics[height=8cm,width=7.5cm]{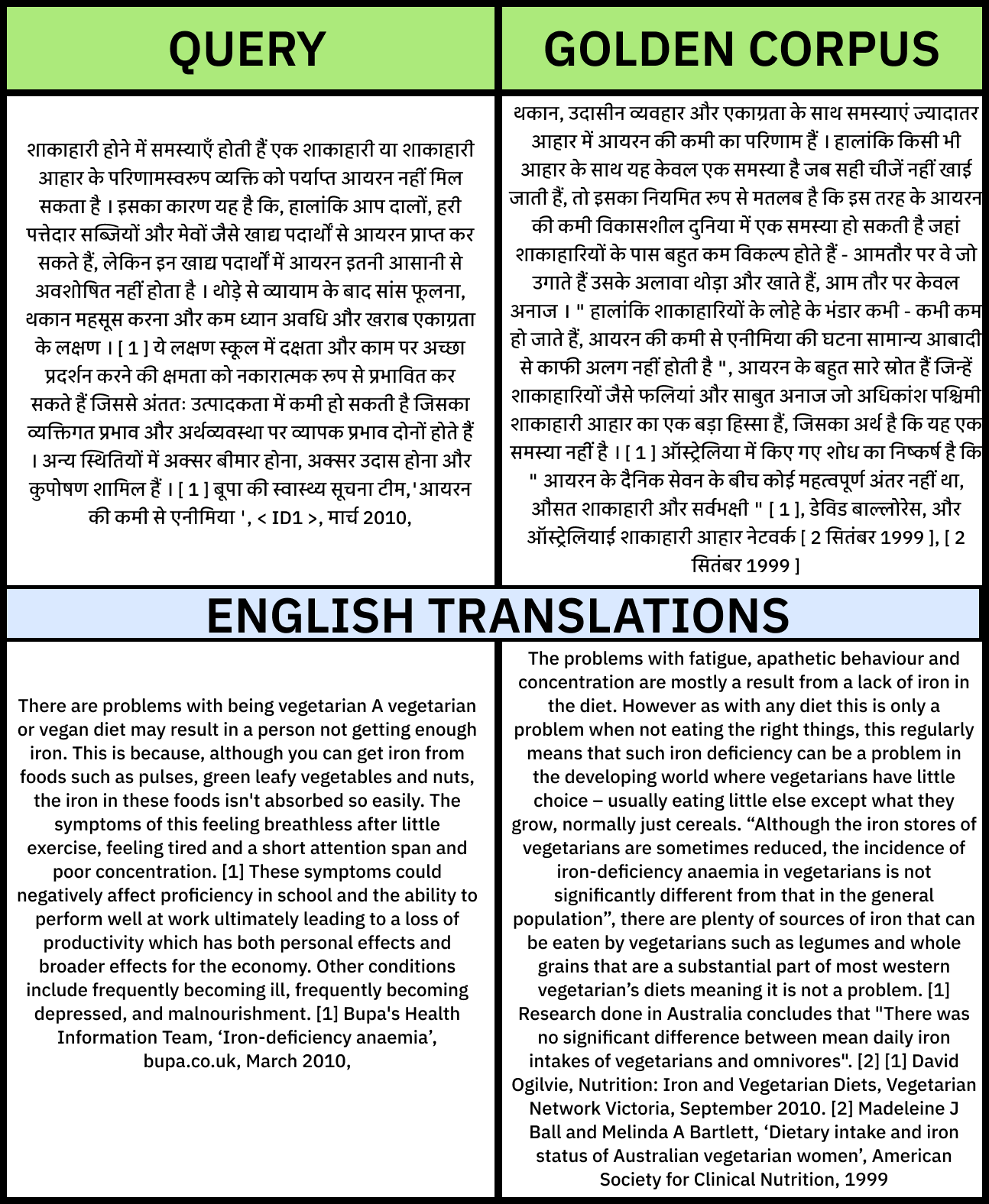}
    \caption{An example of a query with its corresponding golden corpus from the ArguAna Dataset}
    \label{fig:arguana_example}
\end{figure}

An example of a query with its corresponding golden corpus has been provided in Figure \ref{fig:arguana_example}

\begin{figure}[hbt!]
    \centering
    \includegraphics[height=4cm,width=7.5cm]{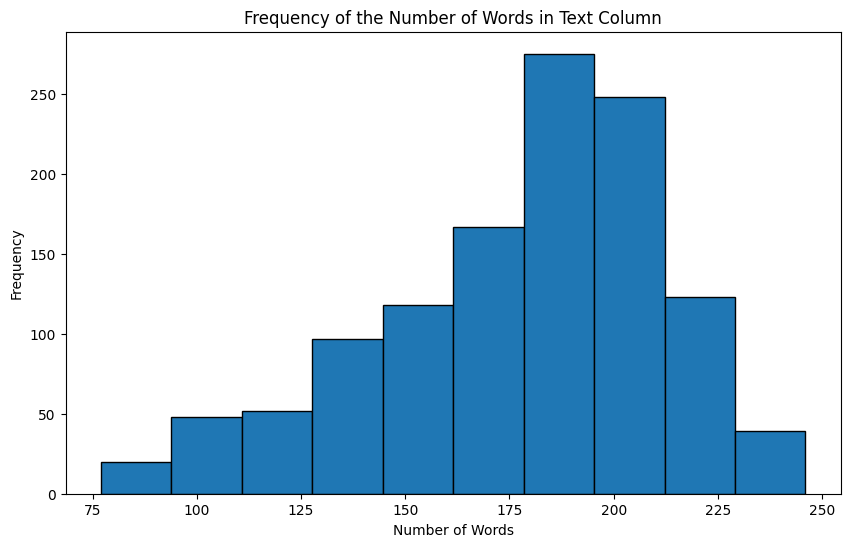}
    \caption{Distribution of the number of words in the queries of ArguAna Dataset}
    \label{fig:arguana_queries}
\end{figure}

\begin{figure}[hbt!]
    \centering
    \includegraphics[height=4cm,width=7.5cm]{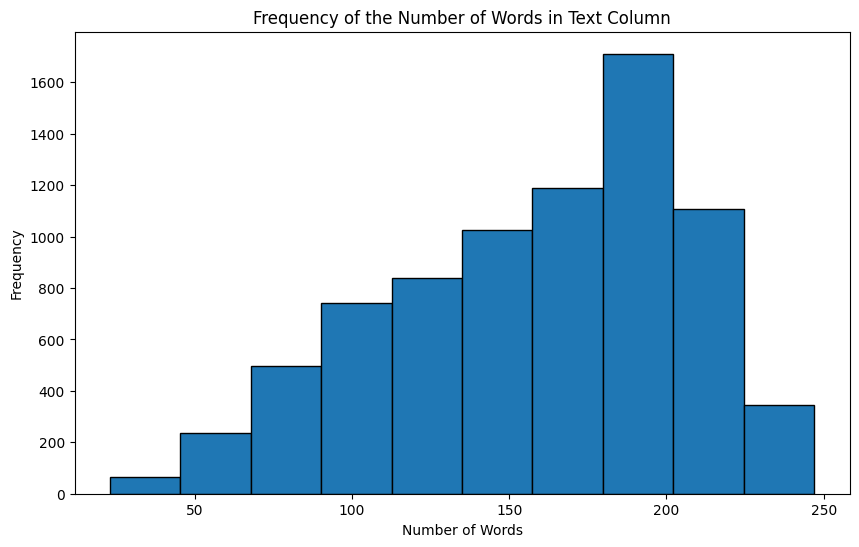}
    \caption{Distribution of Number of Words in the corpus of ArguAna Dataset}
    \label{fig:arguana_corpus}
\end{figure}

Distribution of the number of words in the corpus and queries in the ArguAna dataset has been shown in Figure \ref{fig:arguana_corpus} and Figure \ref{fig:arguana_queries}, respectively.

\subsubsection{FiQA-2018}
\begin{enumerate}
\item \textbf{Task Definition: }It deals with Opinion-Based Question answering. Based on the works of \newcite{10.1145/3184558.3192301},translation of the financial data
extracted by crawling StackExchange posts under the Investment topic from 2009-2017, after passing through filteration processes mentioned in \ref{subsec:translation}, acts as the corpus. While translations from the original training split acts as the queries.

    \item \textbf{Domain :} Finance
\end{enumerate}

\begin{figure}[hbt!]
    \centering
    \includegraphics[height=6cm,width=7.5cm]{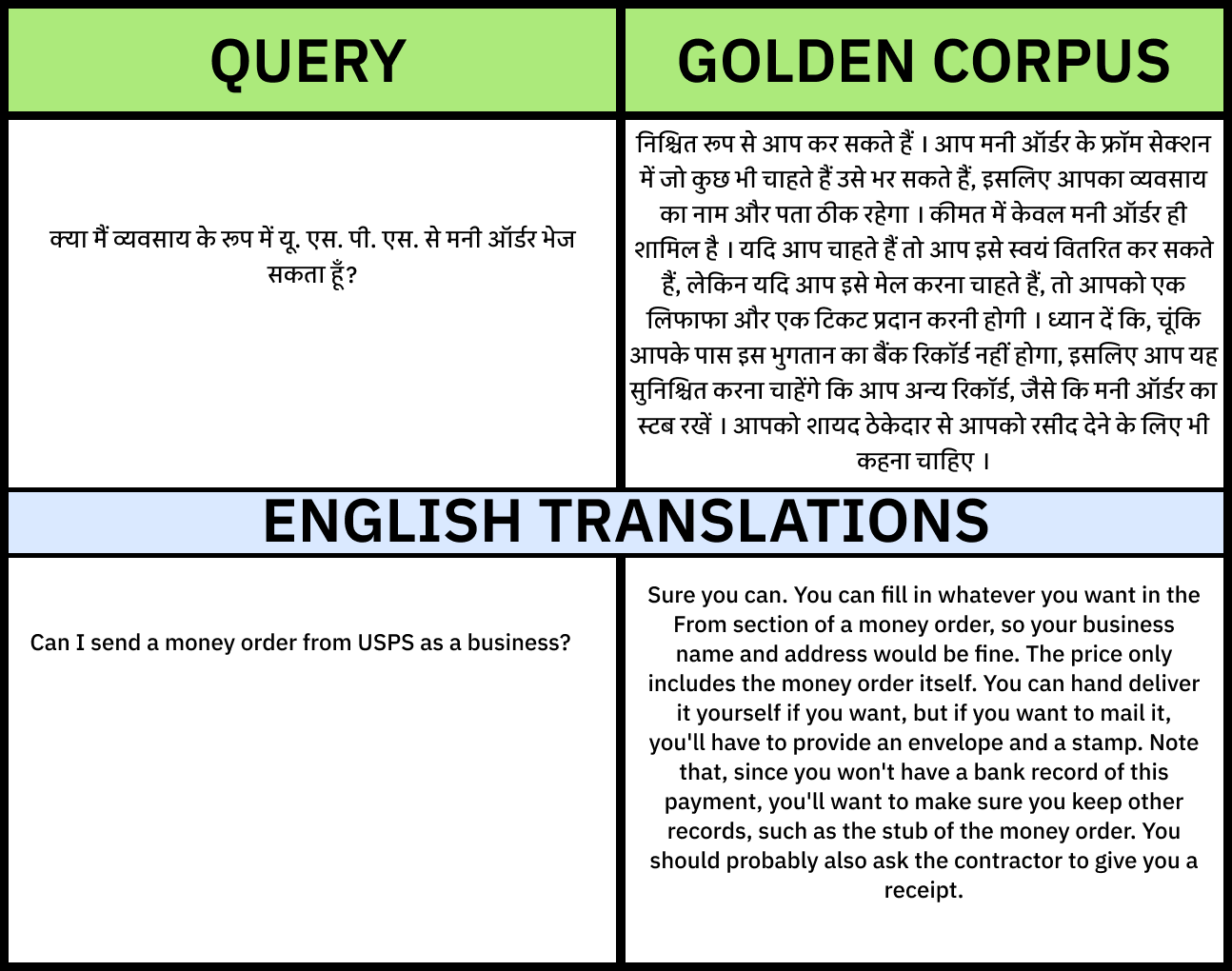}
    \caption{An example of a query with its corresponding golden corpus from the FiQA-2018 Dataset}
    \label{fig:fiqa_example}
\end{figure}

An example of query with its corresponding golden corpus from the FiQA-2018 dataset has been provided in Figure \ref{fig:fiqa_example}

\begin{figure}[hbt!]
    \centering
    \includegraphics[height=4cm,width=7.5cm]{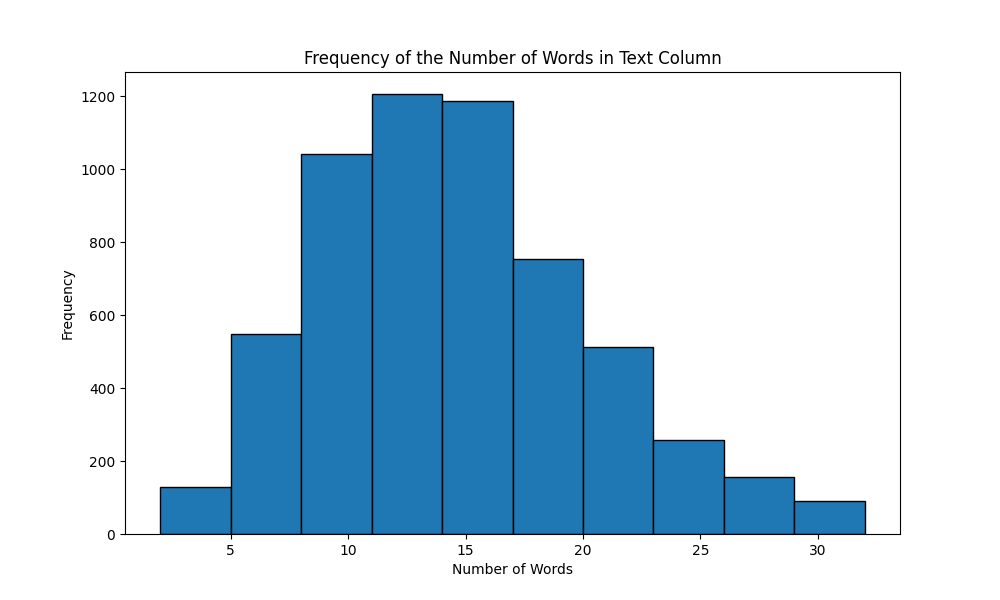}
    \caption{Distribution of the number of words in the queries of FiQA-2018 Dataset}
    \label{fig:fiqa_queries}
\end{figure}

\begin{figure}[hbt!]
    \centering
    \includegraphics[height=4cm,width=7.5cm]{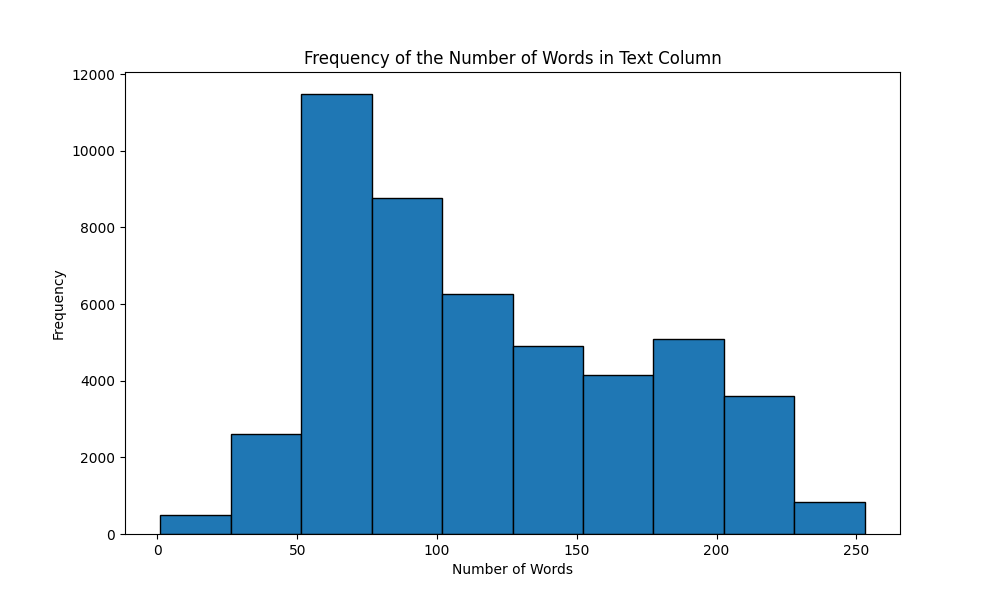}
    \caption{Distribution of Number of Words in corpus of FiQA-2018 Dataset}
    \label{fig:fiqa_corpus}
\end{figure}

Distribution of the number of words in the corpus and queries in FiQA-2018 dataset has been shown in Figure \ref{fig:fiqa_corpus} and Figure \ref{fig:fiqa_queries} respectively.

\subsubsection{TREC-COVID}
\begin{enumerate}
\item \textbf{Task Definition: }\newcite{10.1145/3451964.3451965} introduced the original TREC-COVID dataset which is an ad-hoc seach challenge based on CORD-19 dataset containing articles about the COVID-19 Pandemic. The translated and filtered version of the CORD-19 Dataset constitutes the corpus while the final cumulative judgements with query descriptions from the original task are the queries in the TREC-COVID Dataset.

    \item \textbf{Domain :} Bio-Medical
\end{enumerate}

\begin{figure}[hbt!]
    \centering
    \includegraphics[height=8.5cm,width=7.5cm]{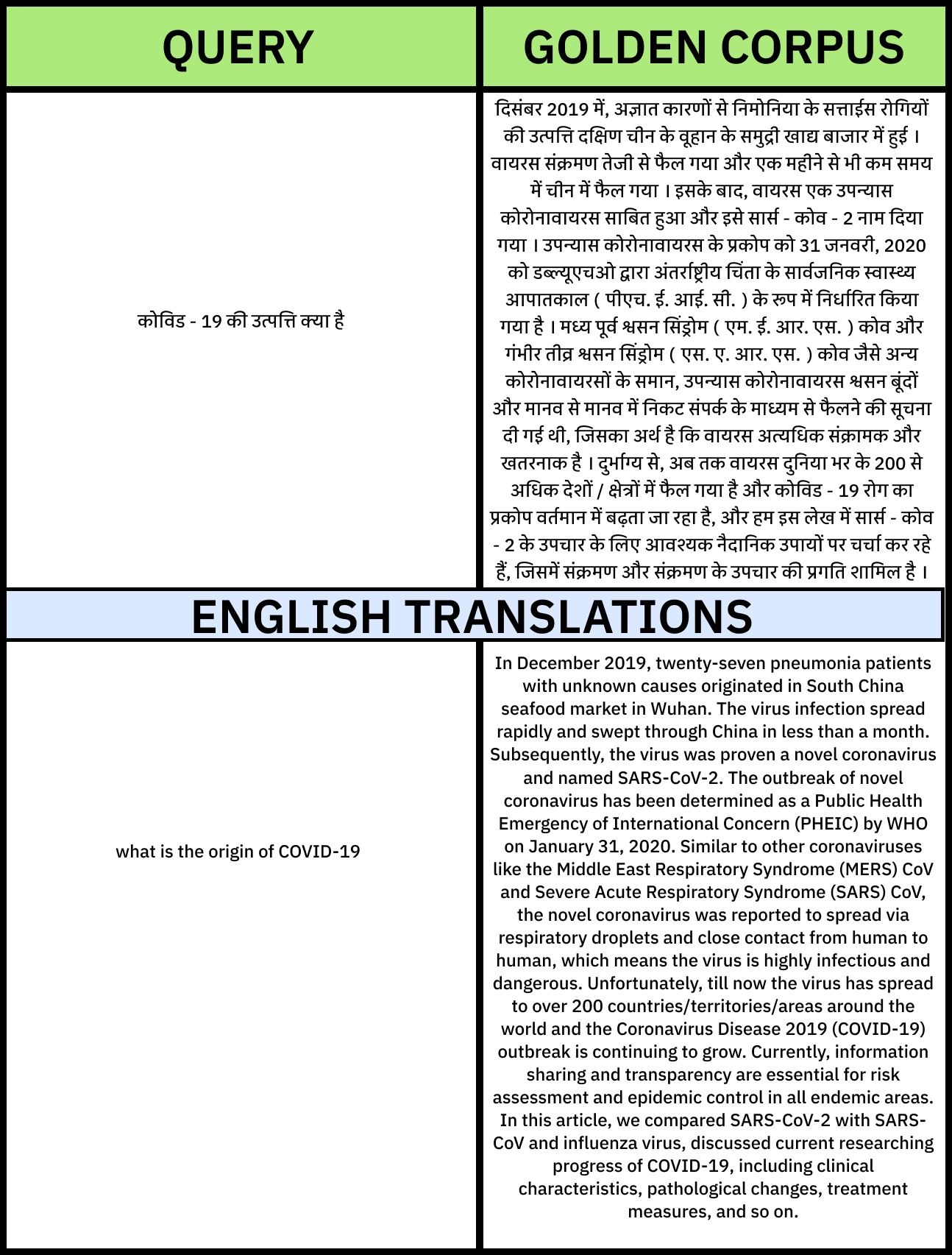}
    \caption{An example of a query with its corresponding golden corpus from the TREC-COVID Dataset}
    \label{fig:trec_example}
\end{figure}

An example of a query with its corresponding golden corpus from the TREC-COVID dataset has been provided in Figure \ref{fig:trec_example}

\begin{figure}[hbt!]
    \centering
    \includegraphics[height=4cm,width=7.5cm]{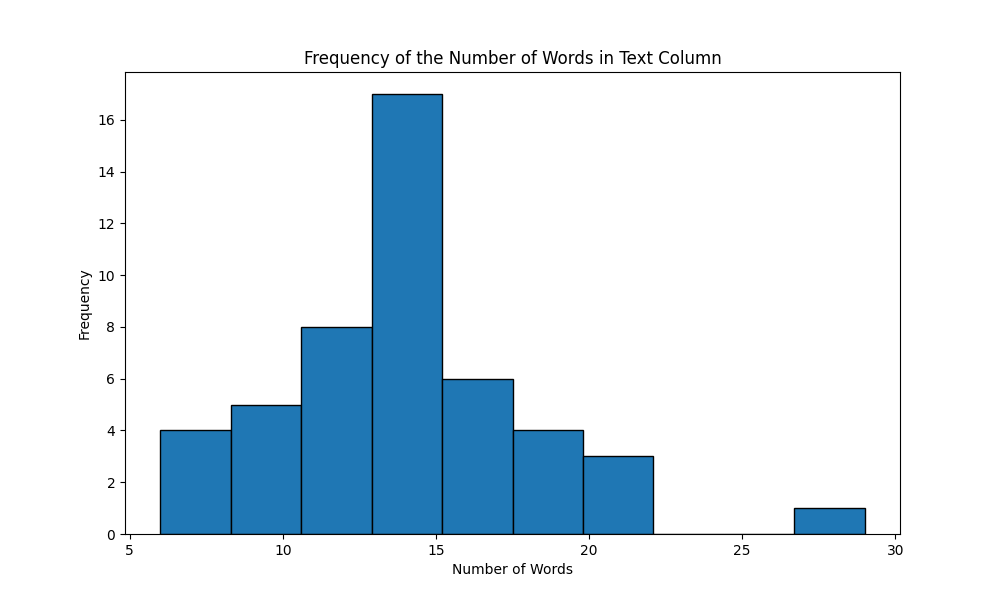}
    \caption{Distribution of the number of words in the queries of TREC-COVID Dataset}
    \label{fig:trec_queries}
\end{figure}

\begin{figure}[hbt!]
    \centering
    \includegraphics[height=4cm,width=7.5cm]{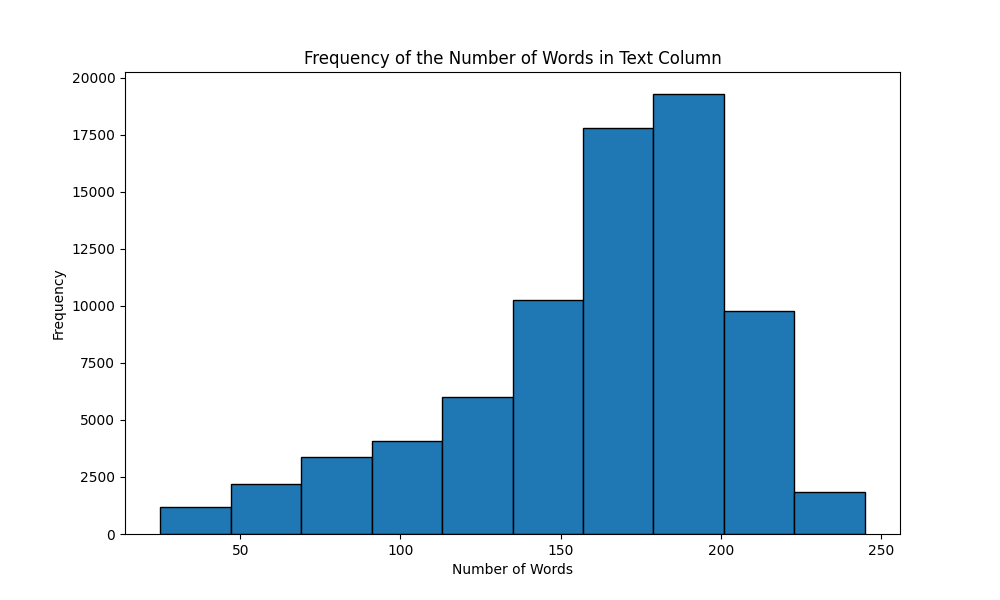}
    \caption{Distribution of Number of Words in the corpus of TREC-COVID Dataset}
    \label{fig:trec_corpus}
\end{figure}

Distribution of the number of words in the corpus and queries in TREC-COVID dataset has been shown in Figure \ref{fig:trec_corpus} and Figure \ref{fig:trec_queries} respectively.

\subsubsection{SCIDOCS}
\begin{enumerate}
\item \textbf{Task Definition: } Inspired by \newcite{cohan-etal-2020-specter}, in this task, we expect the model to retrieve cited papers for a given scientific paper abstract as input. The corpus contains about 22k translated and filtered scientific paper abstracts and 850 translated paper titles as queries.

    \item \textbf{Domain :} Scientific
\end{enumerate} 

\begin{figure}[hbt!]
    \centering
    \includegraphics[height=6cm,width=7.5cm]{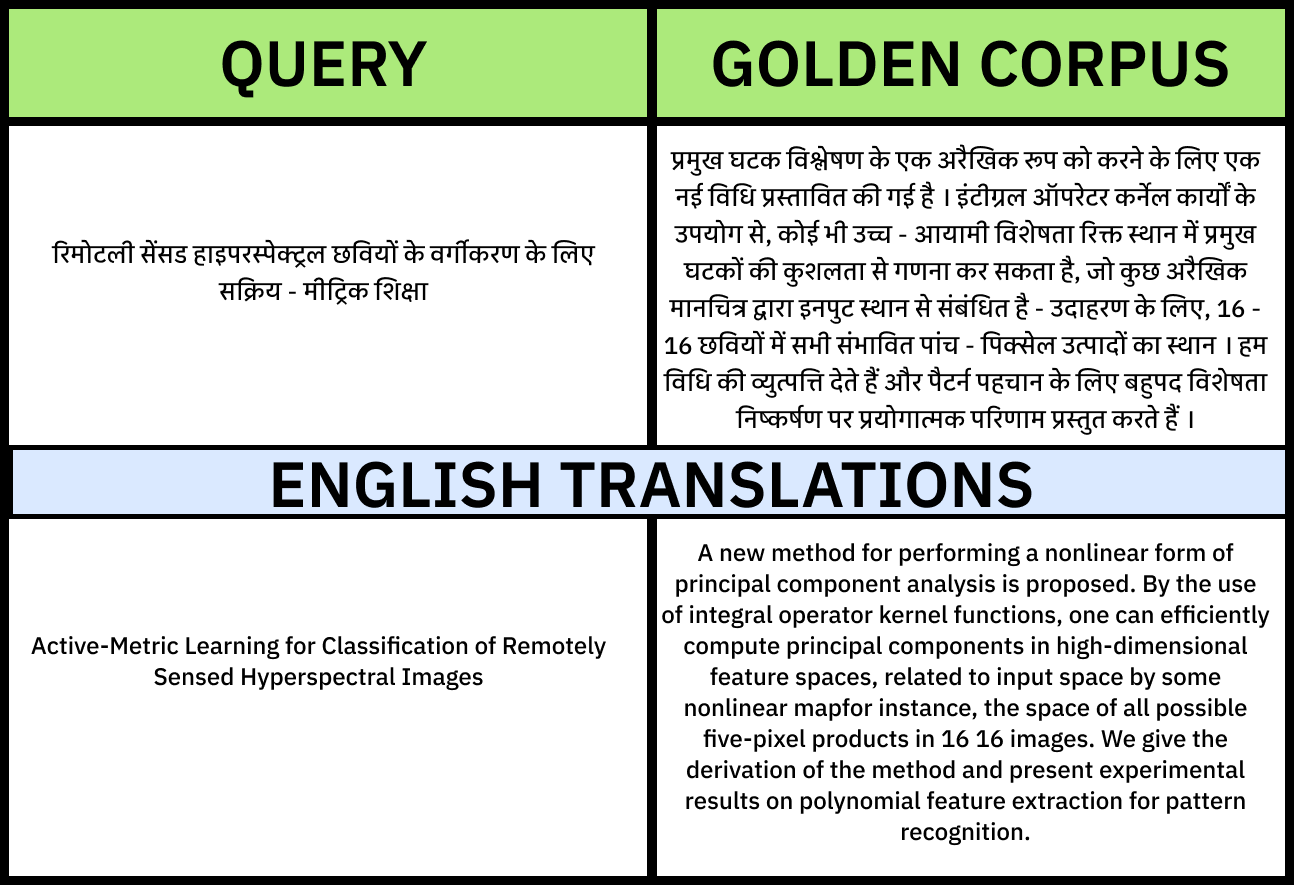}
    \caption{An example of a query with its corresponding golden corpus from the SCIDOCS Dataset}
    \label{fig:scidocs_example}
\end{figure}

An example of query with its corresponding golden corpus from the SCIDOCS dataset has been provided in Figure \ref{fig:scidocs_example}

\begin{figure}[hbt!]
    \centering
    \includegraphics[height=4cm,width=7.5cm]{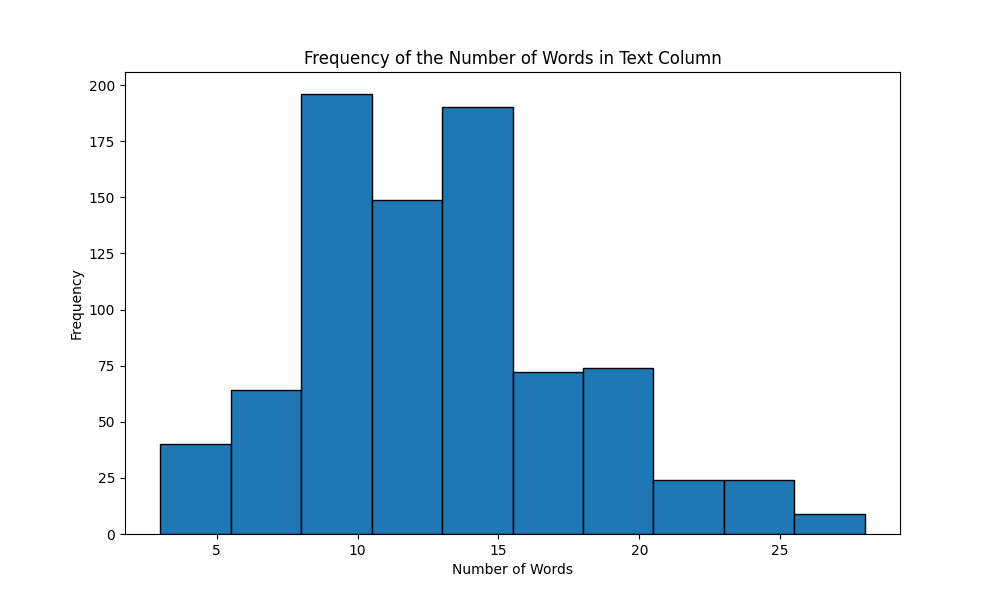}
    \caption{Distribution of the number of words in the queries of SCIDOCS Dataset}
    \label{fig:scidocs_queries}
\end{figure}

\begin{figure}[hbt!]
    \centering
    \includegraphics[height=4cm,width=7.5cm]{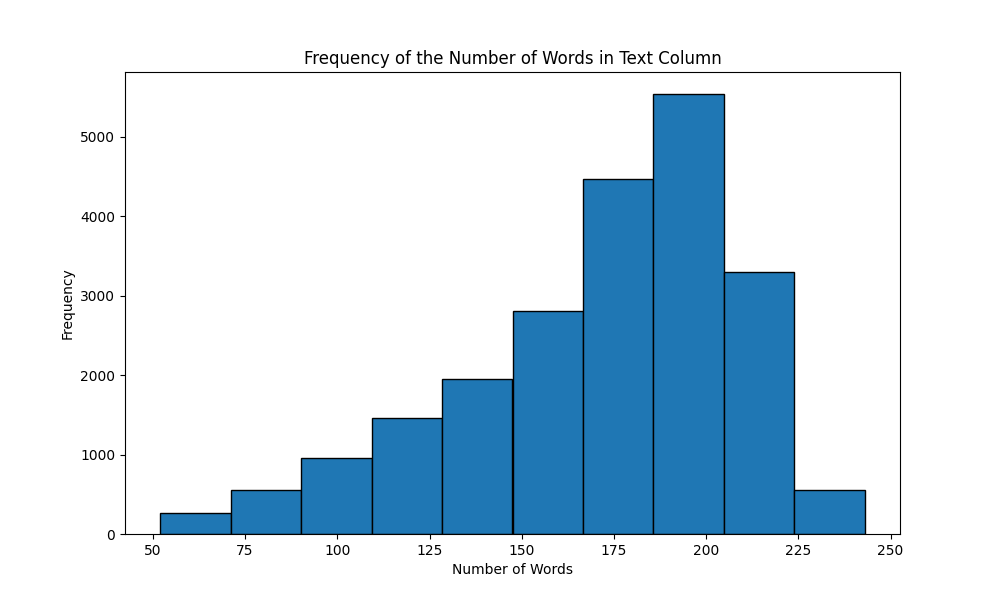}
    \caption{Distribution of Number of Words in the corpus of SCIDOCS Dataset}
    \label{fig:scidocs_corpus}
\end{figure}

Distribution of the number of words in the corpus and queries in the SCIDOCS dataset has been shown in Figure \ref{fig:scidocs_corpus} and Figure \ref{fig:scidocs_queries}, respectively.
\subsubsection{SciFact}
\begin{enumerate}
\item \textbf{Task Definition: } This task involves the verification of scientific claims given the abstract of scientific articles from recent literature. For this task, the model is expected to retrieve relevant abstracts with a given claim as input.

    \item \textbf{Domain :} Scientific
\end{enumerate} 

\begin{figure}[hbt!]
    \centering
    \includegraphics[height=8cm,width=7.5cm]{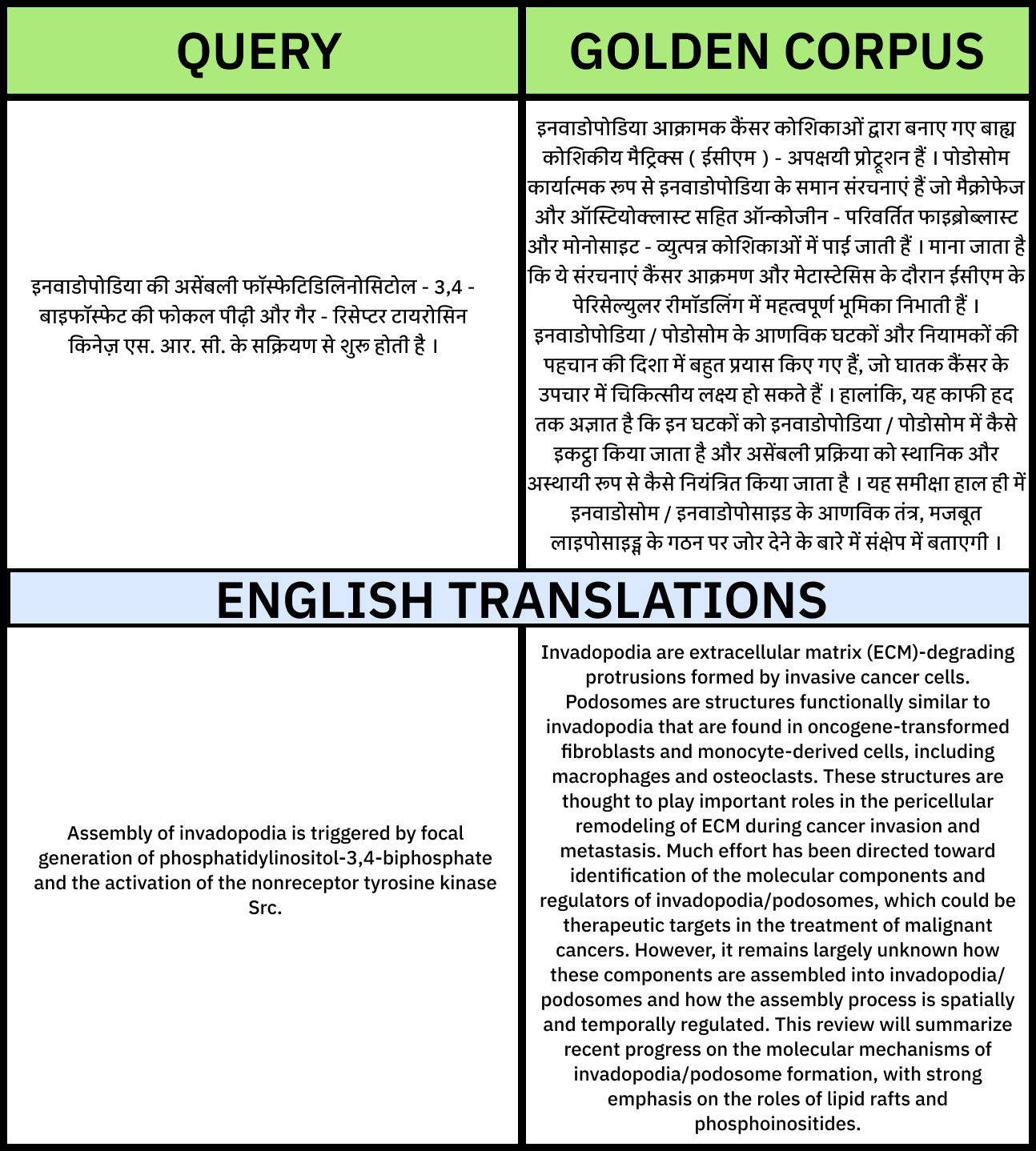}
    \caption{An example of a query with its corresponding golden corpus from the SciFact Dataset}
    \label{fig:scifact_example}
\end{figure}

An example of query with its corresponding golden corpus from the SciFact dataset has been provided in Figure \ref{fig:scifact_example}

\begin{figure}[hbt!]
    \centering
    \includegraphics[height=4cm,width=7.5cm]{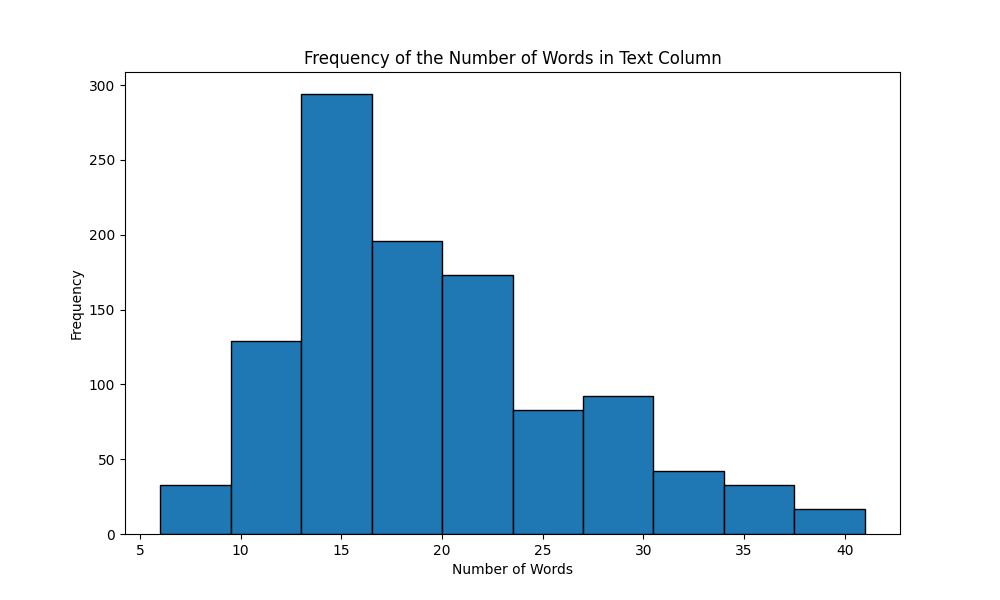}
    \caption{Distribution of the number of words in the queries of SciFact Dataset}
    \label{fig:scifact_queries}
\end{figure}

\begin{figure}[hbt!]
    \centering
    \includegraphics[height=4cm,width=7.5cm]{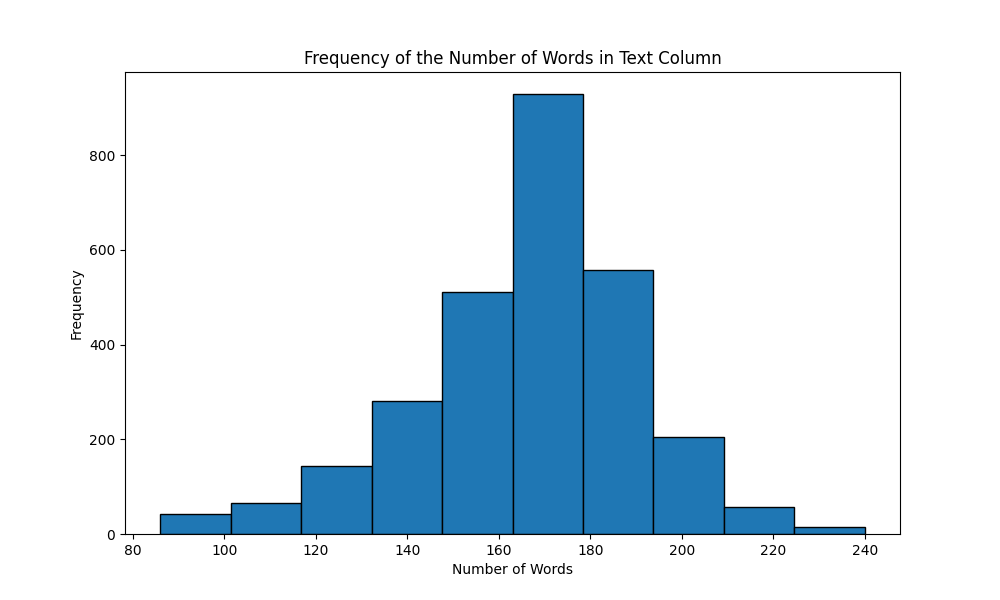}
    \caption{Distribution of Number of Words in the corpus of SciFact Dataset}
    \label{fig:scifact_corpus}
\end{figure}

Distribution of the number of words in the corpus and queries in the SciFact dataset has been shown in Figure \ref{fig:scifact_corpus} and Figure \ref{fig:scifact_queries}, respectively.

\subsubsection{Touch\'e-2020}
\begin{enumerate}
\item \textbf{Task Definition: } Similar to ArguAna this task deals with this task deals with the retrieval of conversational arguments.  The translated and filtered conclusion forms the title and premise for arguments \cite{wachsmuth-etal-2017-building} constitutes the corpus. The translations of the Touch\'e-2020 task data are the queries.

    \item \textbf{Domain :} Miscellaneous
\end{enumerate} 

\begin{figure}[hbt!]
    \centering
    \includegraphics[height=6cm,width=7.5cm]{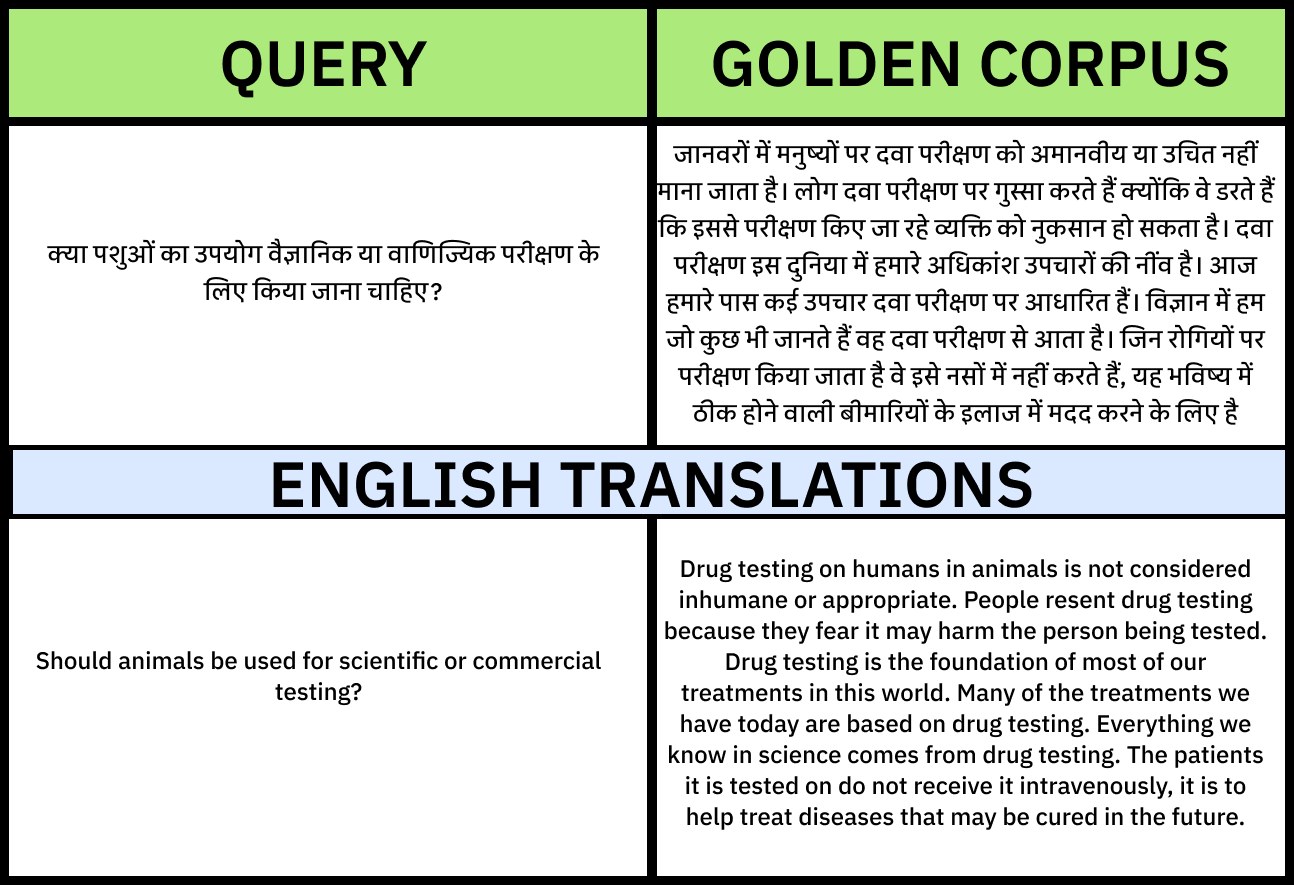}
    \caption{An example of a query with its corresponding golden corpus from the Touch\'e-2020 Dataset}
    \label{fig:touche_example}
\end{figure}

An example of query with its corresponding golden corpus from the Touch\'e-2020 dataset has been provided in Figure \ref{fig:touche_example}

\begin{figure}[hbt!]
    \centering
    \includegraphics[height=4cm,width=7.5cm]{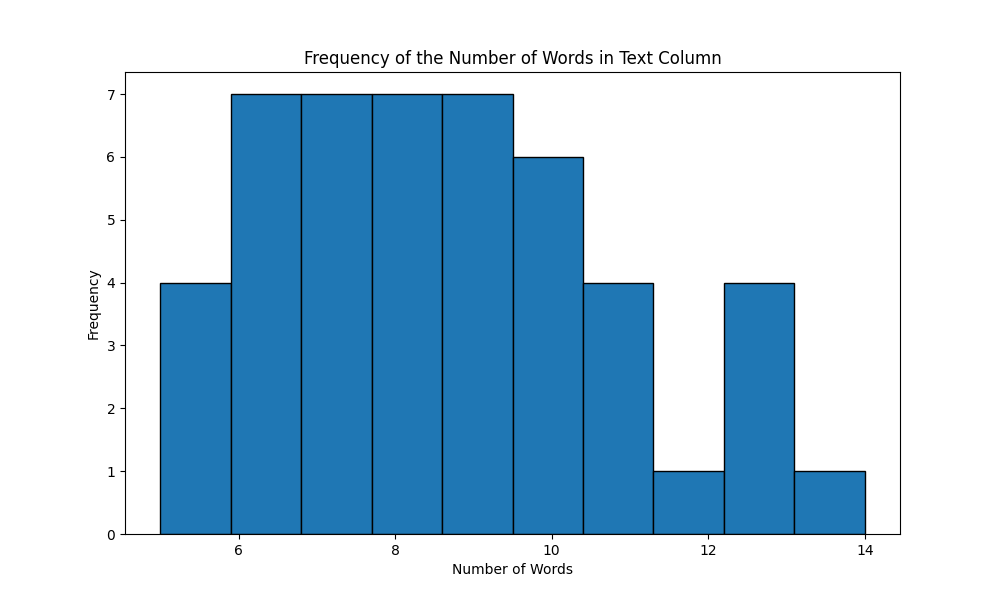}
    \caption{Distribution of the number of words in the queries of Touch\'e-2020 Dataset}
    \label{fig:touche_queries}
\end{figure}

\begin{figure}[hbt!]
    \centering
    \includegraphics[height=4cm,width=7.5cm]{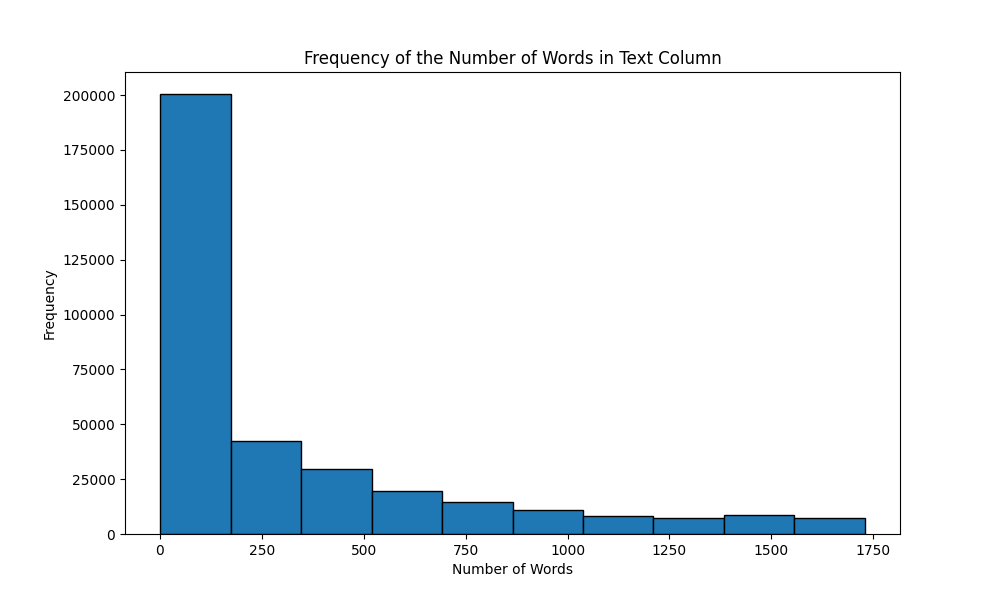}
    \caption{Distribution of Number of Words in the corpus of Touch\'e-2020 Dataset}
    \label{fig:touche_corpus}
\end{figure}

Distribution of the number of words in the corpus and queries in the Touch\'e-2020 dataset has been shown in Figure \ref{fig:touche_corpus} and Figure \ref{fig:touche_queries} respectively.

\subsubsection{NQ}
\begin{enumerate}
\item \textbf{Task Definition: } The task here is to retrieve the correct answer to a given question from a broad corpus of candidate answers. The NQ dataset in the Hindi-BEIR Benchmark is the translated and filtered version of the NQ dataset released by \newcite{thakur2021beir}, which contains Google search queries and documents with paragraphs and answer spans within Wikipedia articles as the corpus.

    \item \textbf{Domain :} WikiPedia
\end{enumerate} 

\begin{figure}[hbt!]
    \centering
    \includegraphics[height=6cm,width=7.5cm]{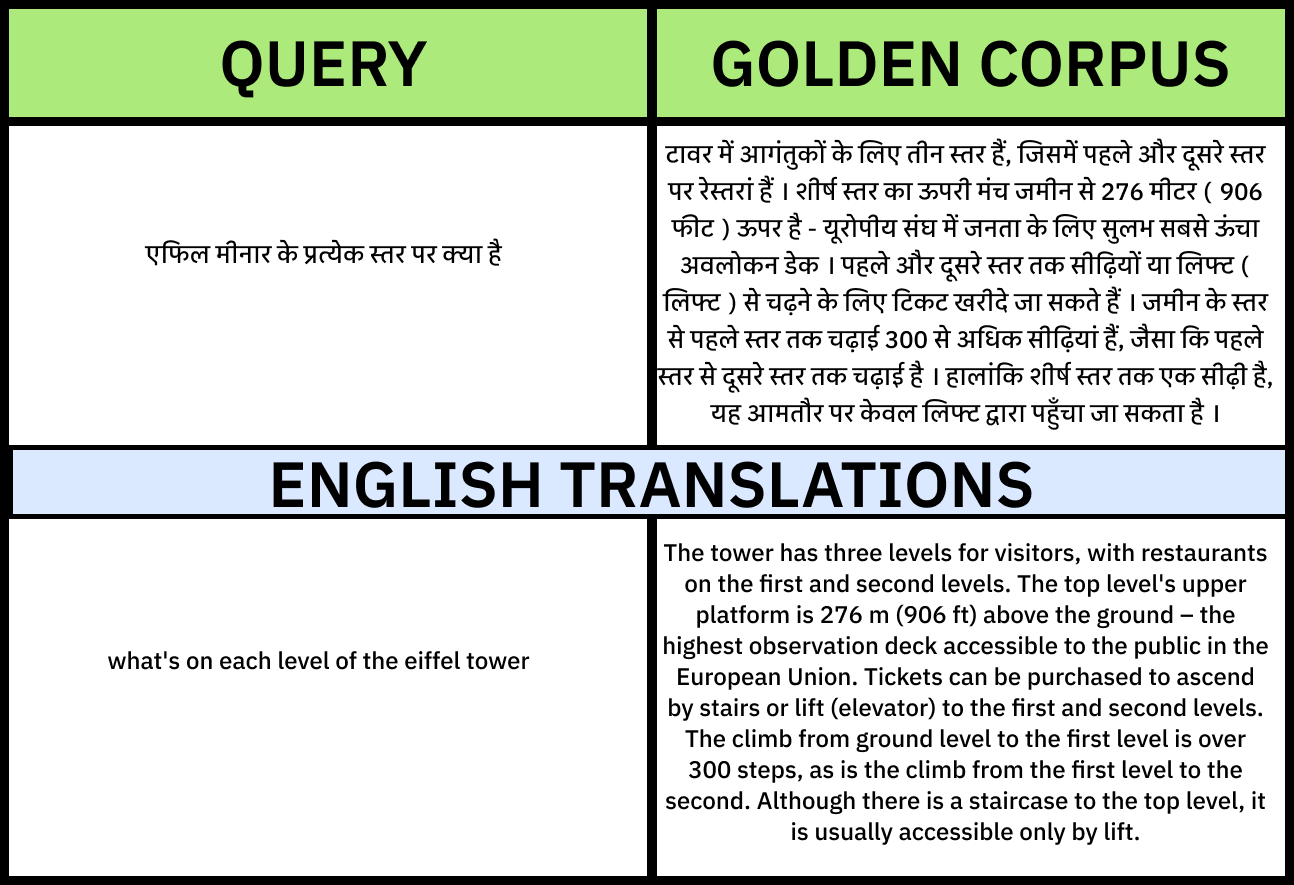}
    \caption{An example of a query with its corresponding golden corpus from the NQ Dataset}
    \label{fig:nq_example}
\end{figure}

An example of a query with its corresponding golden corpus from the NQ dataset has been provided in Figure \ref{fig:nq_example}

\begin{figure}[hbt!]
    \centering
    \includegraphics[height=4cm,width=7.5cm]{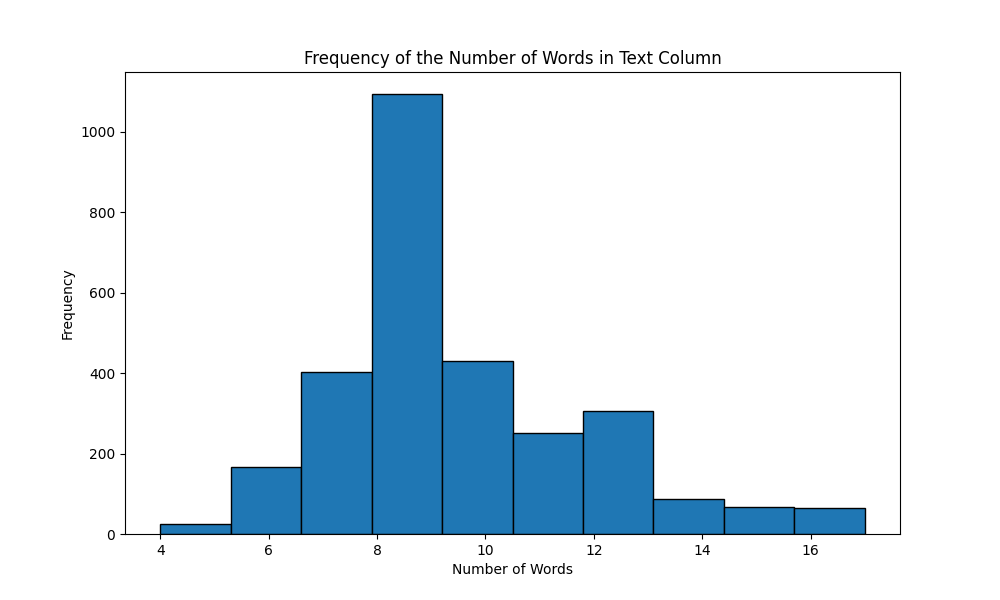}
    \caption{Distribution of the number of words in the queries of NQ Dataset}
    \label{fig:nq_queries}
\end{figure}

\begin{figure}[hbt!]
    \centering
    \includegraphics[height=4cm,width=7.5cm]{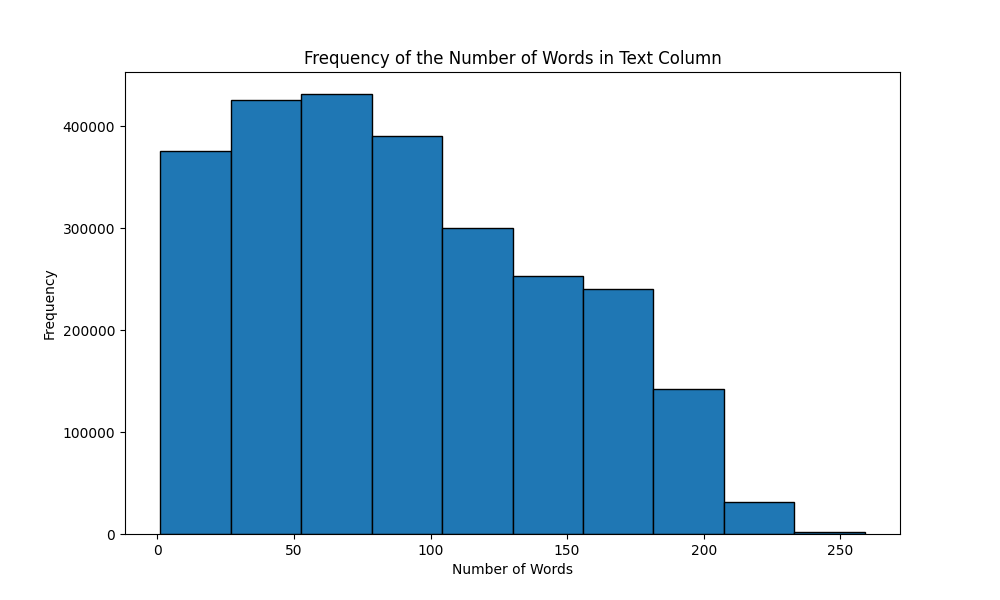}
    \caption{Distribution of Number of Words in the corpus of NQ Dataset}
    \label{fig:nq_corpus}
\end{figure}

Distribution of the number of words in the corpus and queries in the NQ dataset has been shown in Figure \ref{fig:nq_corpus} and Figure \ref{fig:nq_queries}, respectively.

\subsubsection{FEVER}
\begin{figure}[hbt!]
    \centering
    \includegraphics[height=6cm,width=7.5cm]{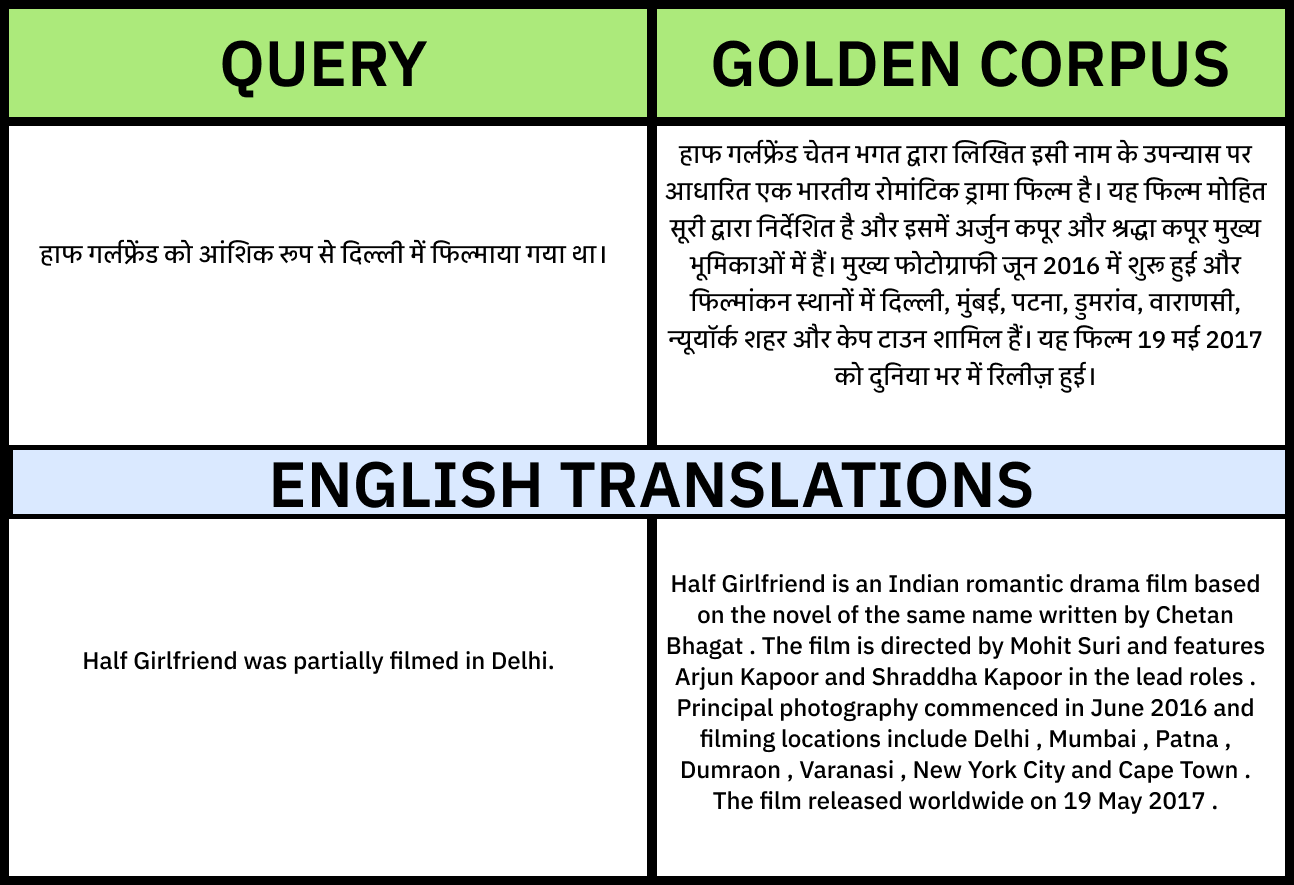}
    \caption{An example of a query with its corresponding golden corpus from the FEVER Dataset}
    \label{fig:fever_example}
\end{figure}

\begin{enumerate}

\item \textbf{Task Definition: } Similar to SciFact, here, the task is to retrieve relevant documents that claim a given fact (query). We translate and filter the test split of the FEVER dataset as proposed by \newcite{thakur2021beir} and include it in the Hindi-BEIR Benchmark.
    \item \textbf{Domain :} Wikipedia
\end{enumerate}
An example of a query with its corresponding golden corpus from the FEVER dataset has been provided in Figure \ref{fig:fever_example}

\begin{figure}[hbt!]
    \centering
    \includegraphics[height=4cm,width=7.5cm]{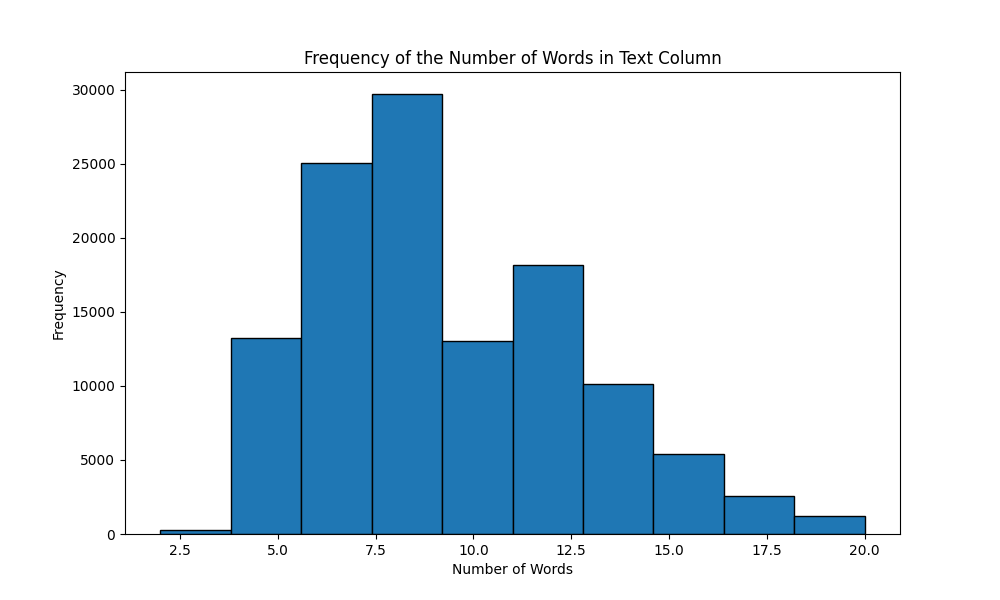}
    \caption{Distribution of the number of words in the queries of FEVER Dataset}
    \label{fig:fever_queries}
\end{figure}

\begin{figure}[hbt!]
    \centering
    \includegraphics[height=4cm,width=7.5cm]{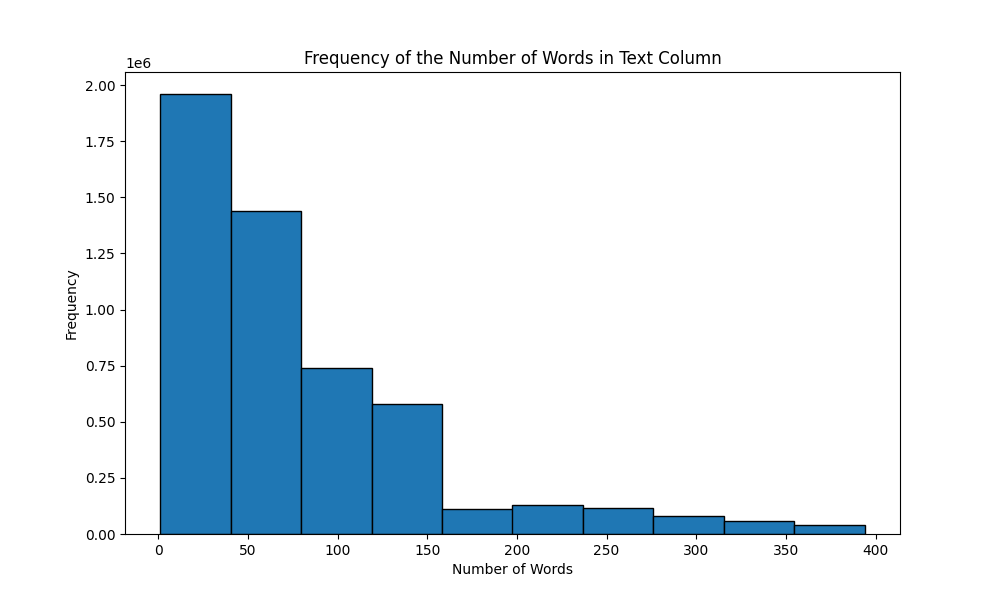}
    \caption{Distribution of Number of Words in the corpus of FEVER Dataset}
    \label{fig:fever_corpus}
\end{figure}

Distribution of the number of words in the corpus and queries in the FEVER dataset has been shown in Figure \ref{fig:fever_corpus} and Figure \ref{fig:fever_queries}, respectively.

\subsubsection{Climate-FEVER}
\begin{figure}[hbt!]
    \centering
    \includegraphics[height=8cm,width=7.5cm]{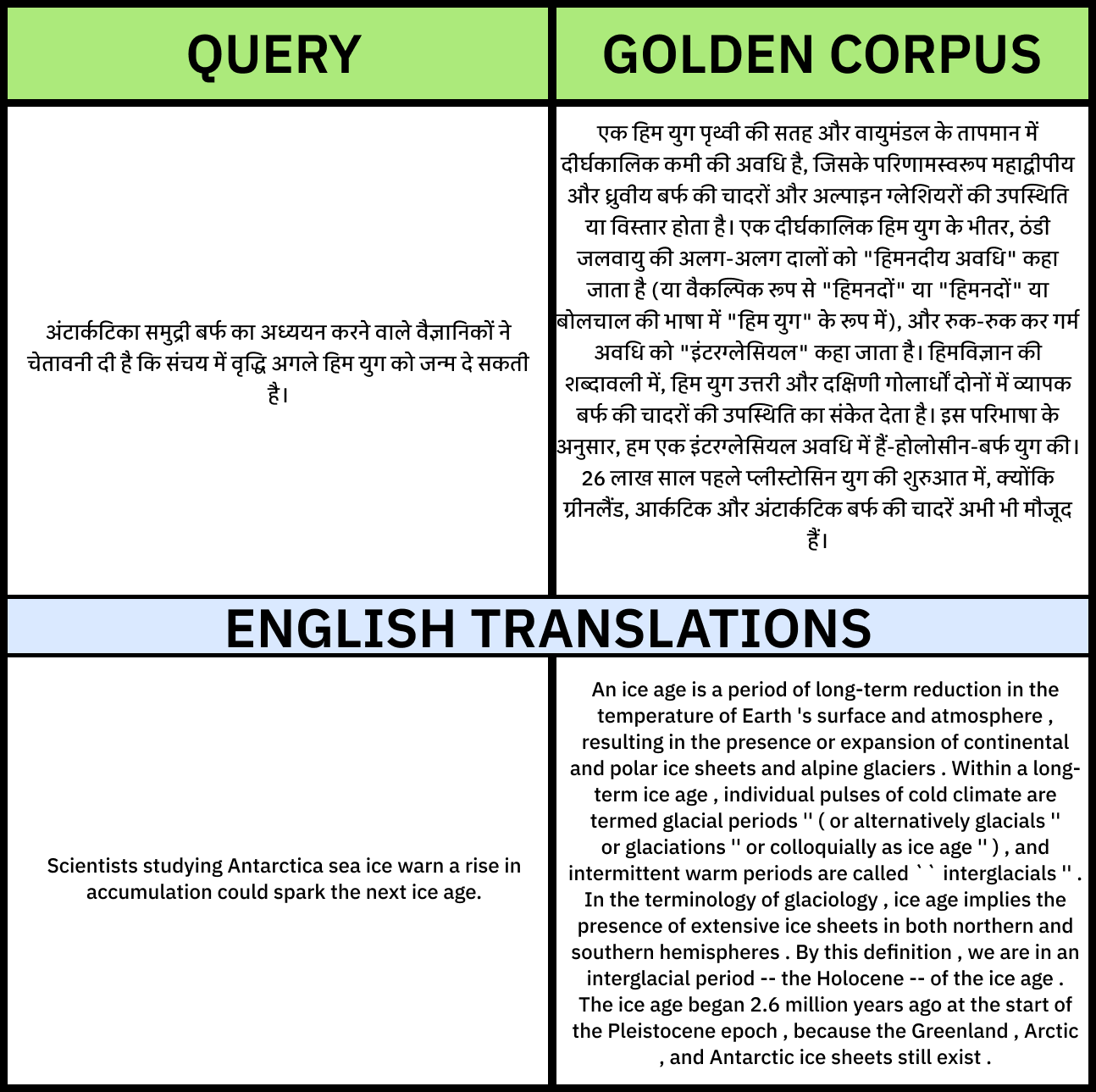}
    \caption{An example of a query with its corresponding golden corpus from the Climate-FEVER Dataset}
    \label{fig:climate_fever_example}
\end{figure}

\begin{enumerate}

\item \textbf{Task Definition: } Similar to FEVER, Climate-FEVER is a dataset for the verification of real-world climate claims. We translate and filter the test split of the Climate-FEVER dataset as proposed by \newcite{thakur2021beir} and include it in the Hindi-BEIR Benchmark.
    \item \textbf{Domain :} Wikipedia
\end{enumerate}
An example of query with its corresponding golden corpus from the Climate-FEVER dataset has been provided in Figure \ref{fig:climate_fever_example}

\begin{figure}[hbt!]
    \centering
    \includegraphics[height=4cm,width=7.5cm]{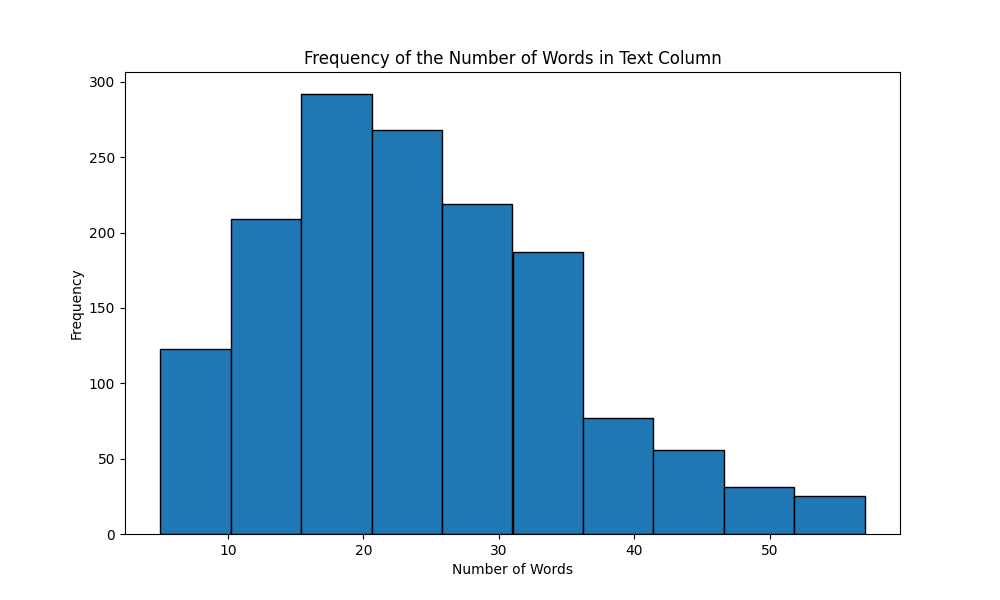}
    \caption{Distribution of the number of words in the queries of Climate-FEVER Dataset}
    \label{fig:climate_fever_queries}
\end{figure}

\begin{figure}[hbt!]
    \centering
    \includegraphics[height=4cm,width=7.5cm]{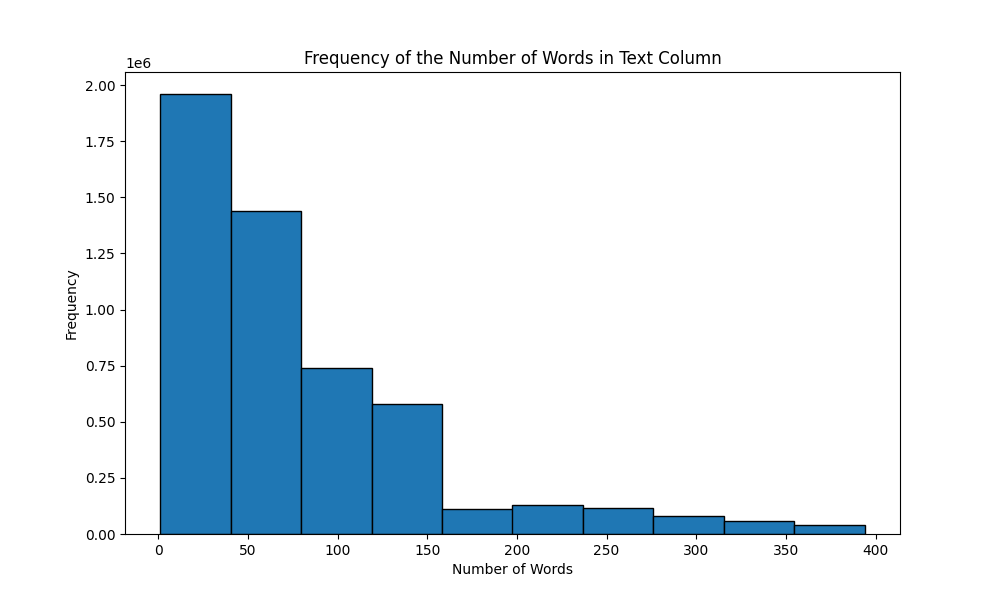}
    \caption{Distribution of Number of Words in corpus of Climate-FEVER Dataset}
    \label{fig:climate_fever_corpus}
\end{figure}

Distribution of the number of words in the corpus and queries in the Climate-FEVER dataset has been shown in Figure \ref{fig:climate_fever_corpus} and Figure \ref{fig:climate_fever_queries}, respectively.

\subsubsection{CC News Retrieval}

\begin{figure}[hbt!]
    \centering
    \includegraphics[height=8.5cm,width=7.5cm]{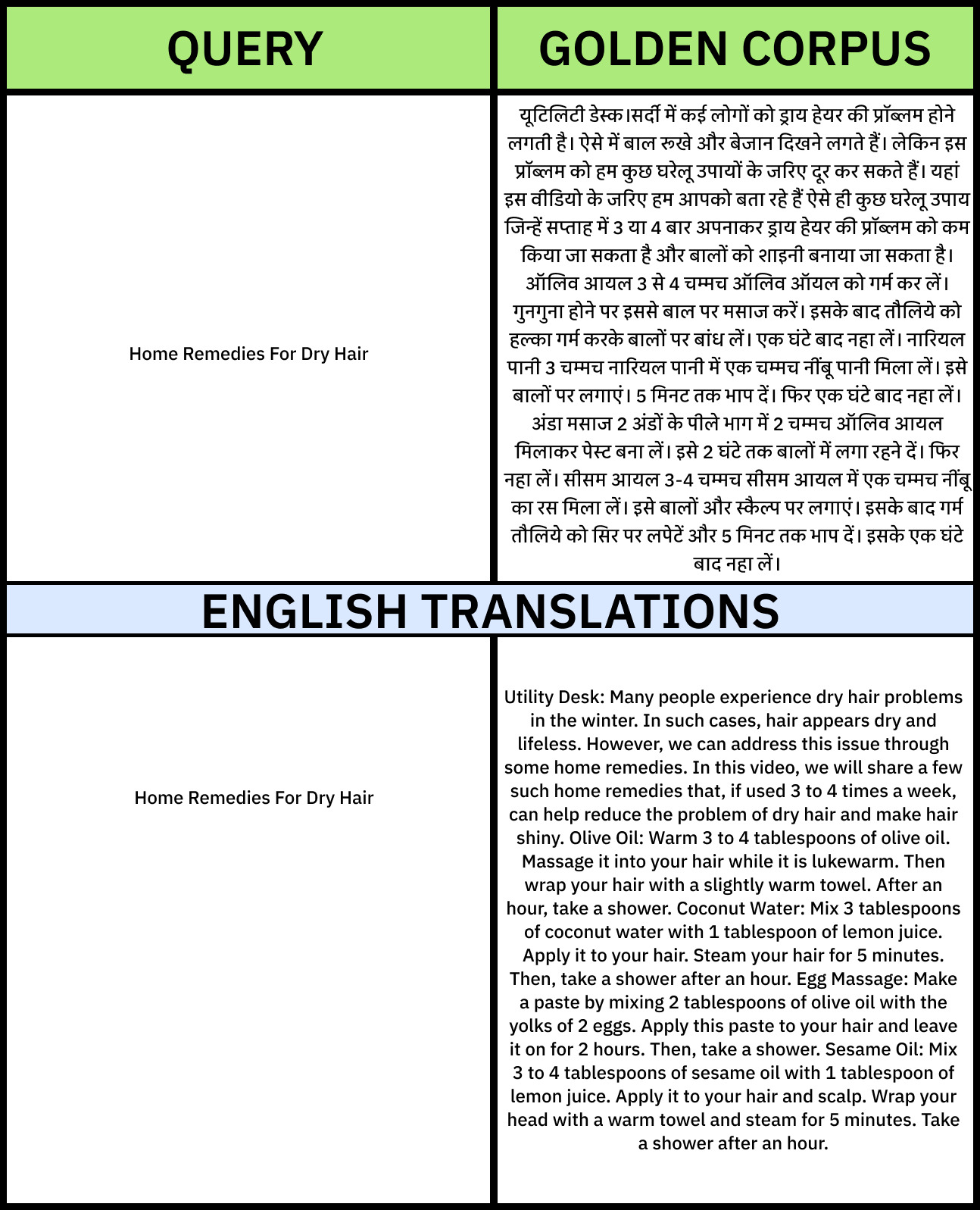}
    \caption{An example of a query with its corresponding golden corpus from the CC News Retrieval Dataset}
    \label{fig:ccnews_example}
\end{figure}

\begin{enumerate}

\item \textbf{Task Definition: } CC News Retrieval Introduces the task to retrieving relevant news articles given a news title. Subsection \ref{cc_news} talks in detail about the data creation process.
    \item \textbf{Domain :} News
\end{enumerate}
An example of a query with its corresponding golden corpus from the CC News Retrieval dataset has been provided in Figure \ref{fig:ccnews_example}

\begin{figure}[hbt!]
    \centering
    \includegraphics[height=4cm,width=7.5cm]{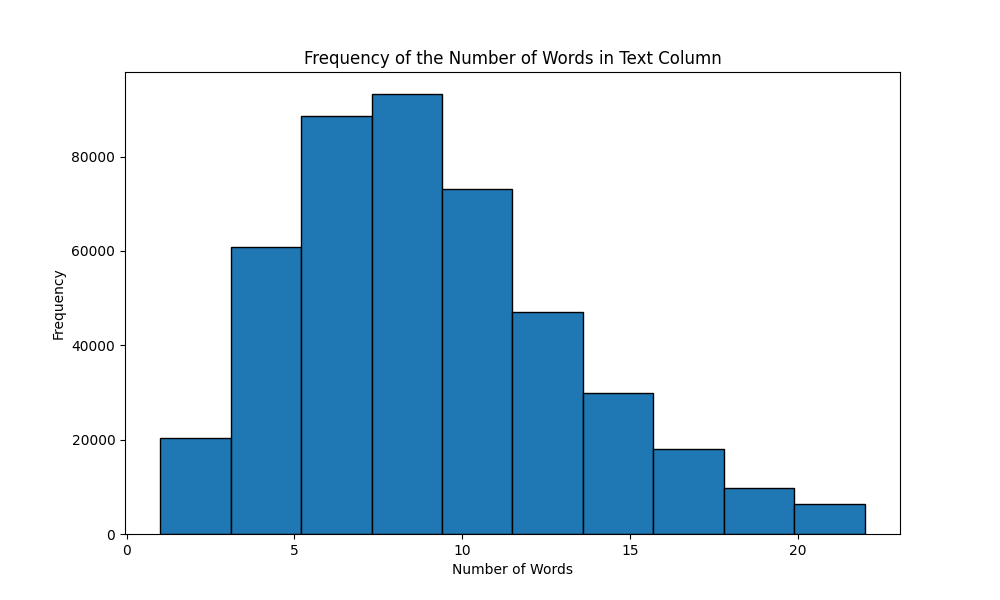}
    \caption{Distribution of the number of words in the queries of CC News Retrieval Dataset}
    \label{fig:ccnews_queries}
\end{figure}

\begin{figure}[hbt!]
    \centering
    \includegraphics[height=4cm,width=7.5cm]{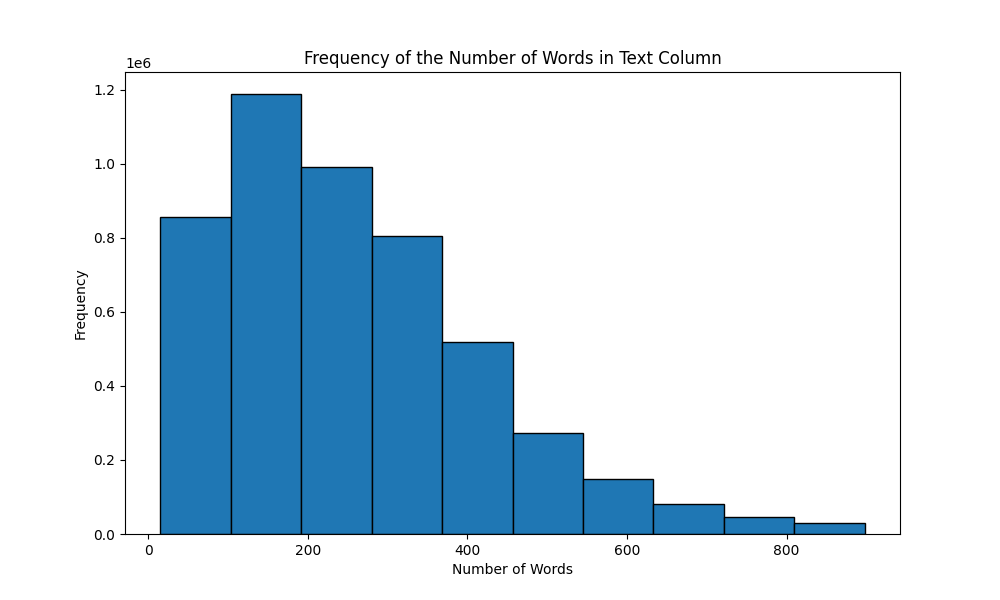}
    \caption{Distribution of Number of Words in corpus of CC News Retrieval Dataset}
    \label{fig:ccnews_corpus}
\end{figure}

Distribution of the number of words in the corpus and queries in CC News Retrieval dataset has been shown in Figure \ref{fig:ccnews_corpus} and Figure \ref{fig:ccnews_queries} respectively.

\subsubsection{Sangraha-IR}

\begin{figure}[hbt!]
    \centering
    \includegraphics[height=8.5cm,width=7.5cm]{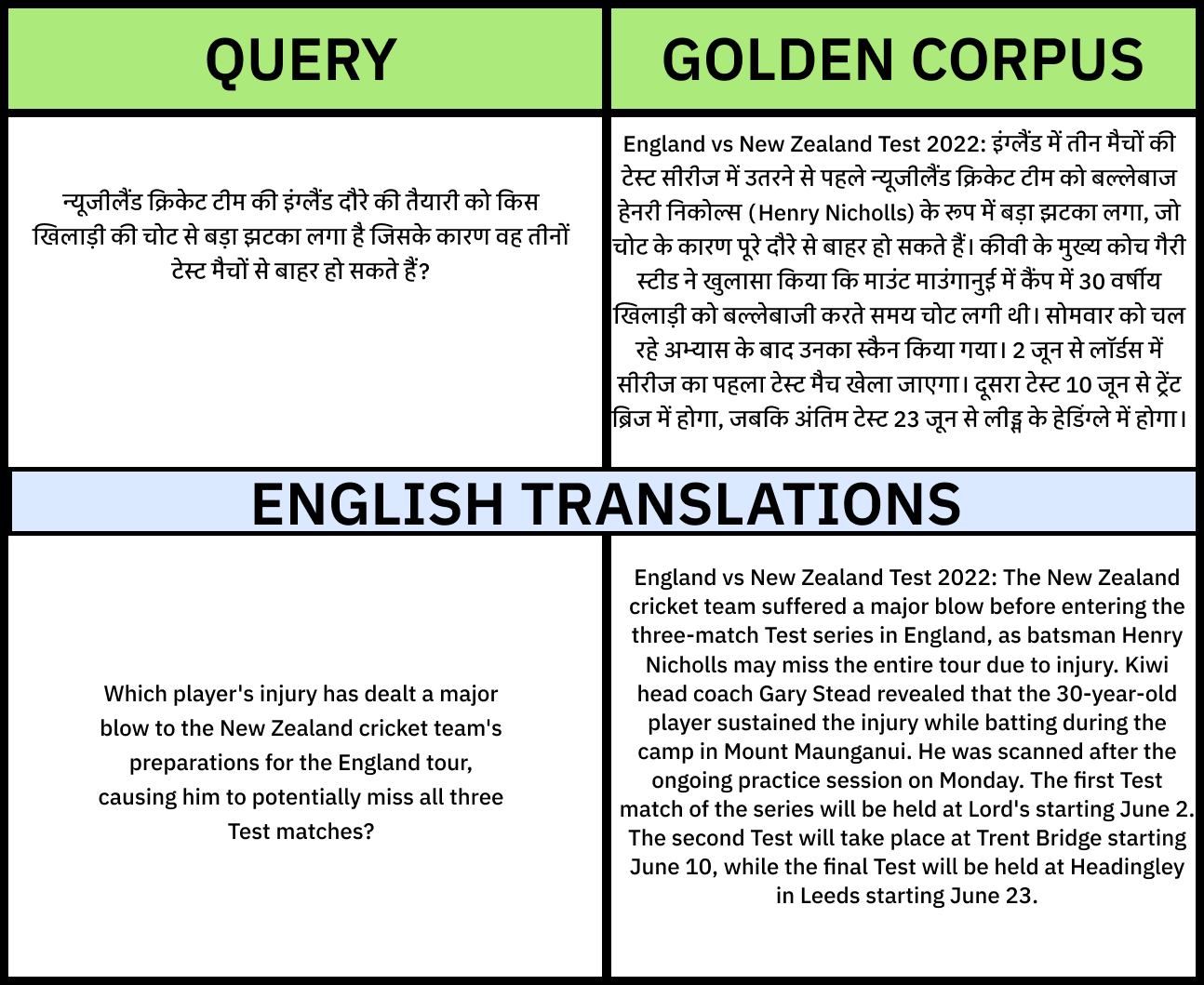}
    \caption{An example of a query with its corresponding golden corpus from the Sangraha-IR Dataset}
    \label{fig:sangraha_example}
\end{figure}

\begin{enumerate}

\item \textbf{Task Definition: } Sangraha-IR introduces the task to retrieving relevant text documents given a question in Hindi. Subsection \ref{cc_news} talks in detail about the data creation process.
    \item \textbf{Domain :} Miscellaneous
\end{enumerate}
An example of a query with its corresponding golden corpus from the Sangraha-IR dataset has been provided in Figure \ref{fig:sangraha_example}

\begin{figure}[hbt!]
    \centering
    \includegraphics[height=4cm,width=7.5cm]{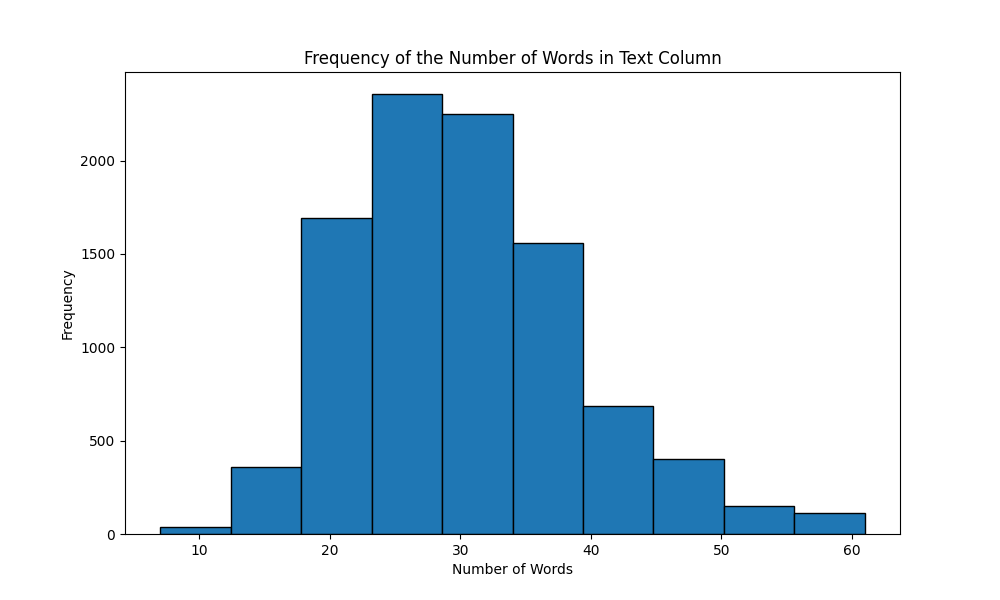}
    \caption{Distribution of the number of words in the queries of Sangraha-IR Dataset}
    \label{fig:sangraha_queries}
\end{figure}

\begin{figure}[hbt!]
    \centering
    \includegraphics[height=4cm,width=7.5cm]{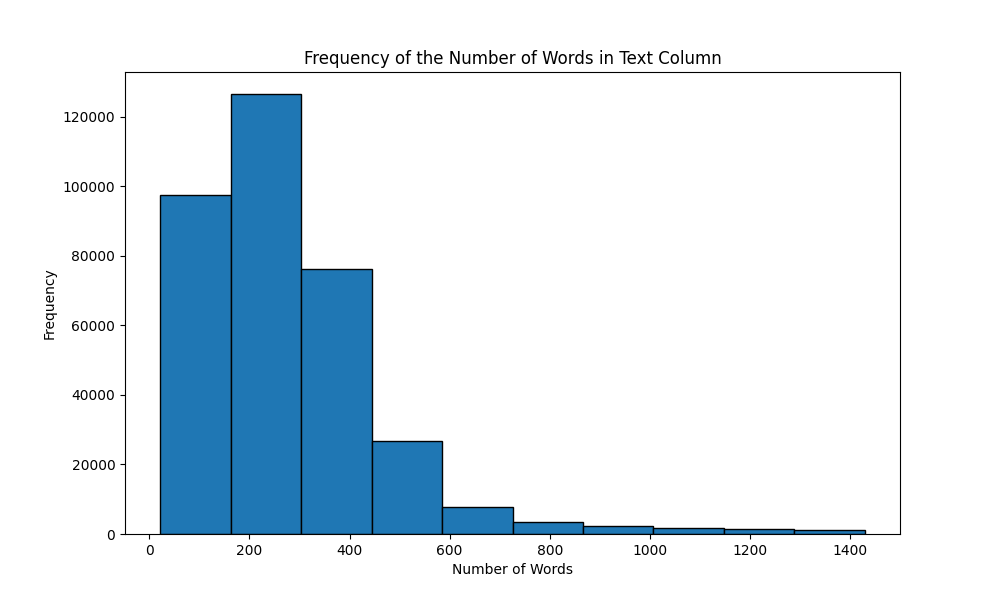}
    \caption{Distribution of Number of Words in corpus of Sangraha-IR Dataset}
    \label{fig:sangraha_corpus}
\end{figure}

Distribution of the number of words in the corpus and queries in the Sangraha-IR dataset has been shown in Figure \ref{fig:sangraha_corpus} and Figure \ref{fig:sangraha_queries}, respectively.





\subsubsection{MIRACL}

\begin{figure}[hbt!]
    \centering
    \includegraphics[height=8cm,width=7.5cm]{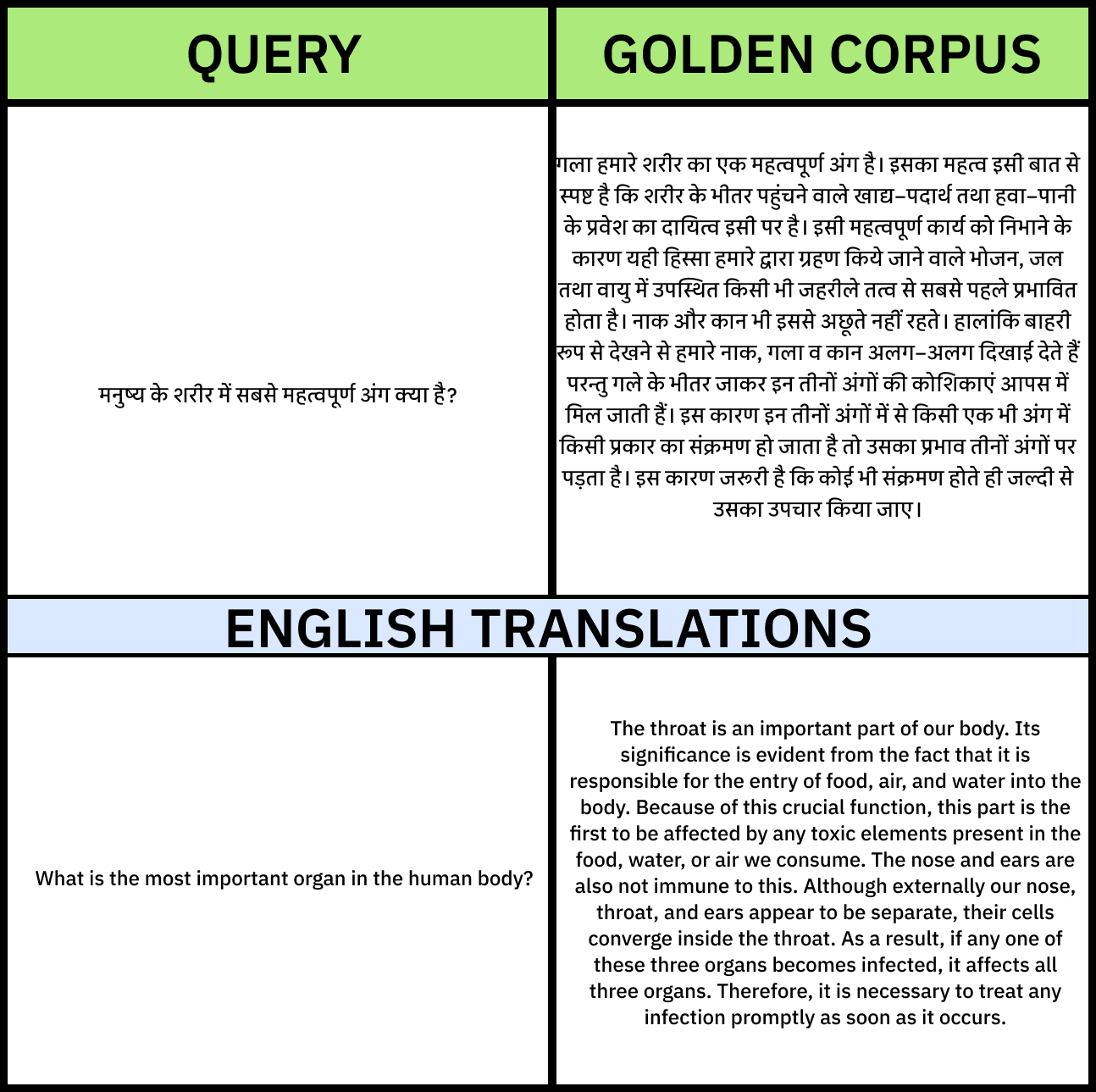}
    \caption{An example of a query with its corresponding golden corpus from the MIRACL Dataset}
    \label{fig:miracl_example}
\end{figure}

\begin{enumerate}

\item \textbf{Task Definition: } MIRACL has been created from the Wikipedia dump of each language. Only plain text is considered, while images, etc., are omitted. The plain text is split up into multiple paragraphs, which act as the corpus. We consider the dev split of the Hindi subset of the original MIRACL in the Hindi-BEIR Benchmark.

    \item \textbf{Domain :} Wikipedia
\end{enumerate}
An example of query with its corresponding golden corpus from the MIRACL dataset has been provided in Figure \ref{fig:miracl_example}

\begin{figure}[hbt!]
    \centering
    \includegraphics[height=4cm,width=7.cm]{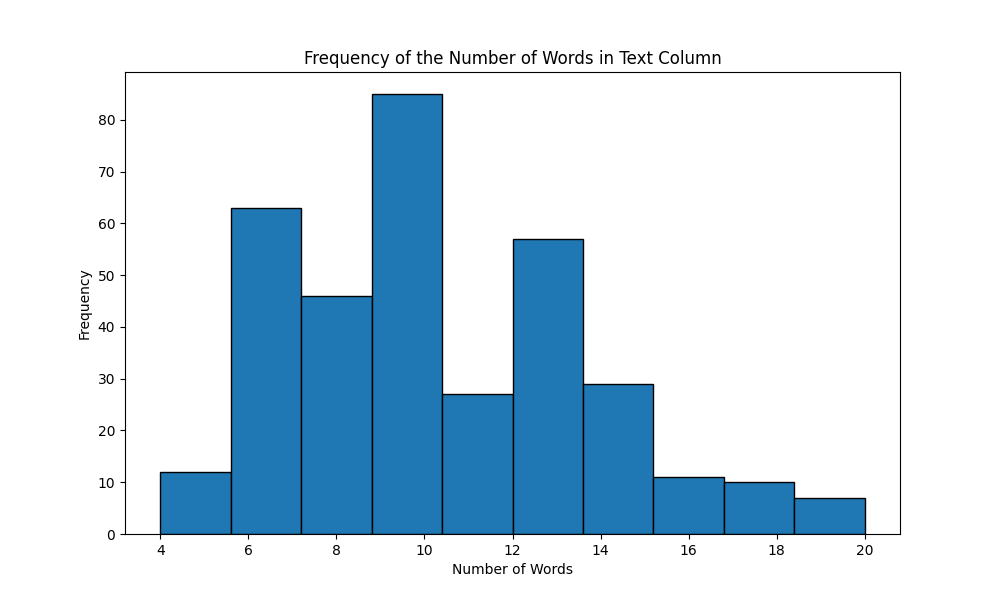}
    \caption{Distribution of the number of words in the queries of MIRACL Dataset}
    \label{fig:miracl_queries}
\end{figure}

\begin{figure}[hbt!]
    \centering
    \includegraphics[height=4cm,width=7.5cm]{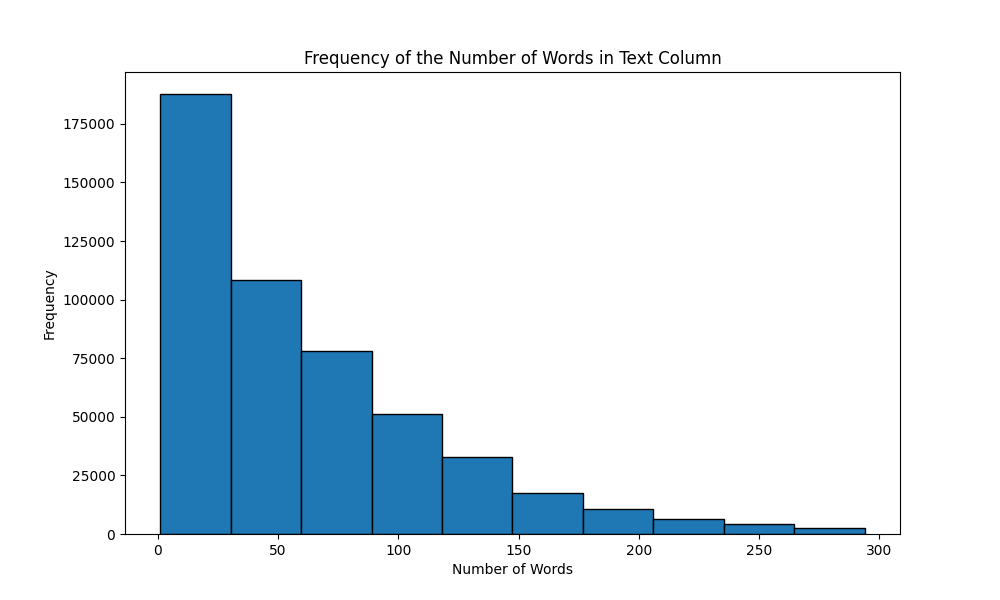}
    \caption{Distribution of Number of Words in the corpus of MIRACL Dataset}
    \label{fig:miracl_corpus}
\end{figure}

Distribution of the number of words in the corpus and queries in the MIRACL dataset has been shown in Figure \ref{fig:miracl_corpus} and Figure \ref{fig:miracl_queries} respectively.

\subsubsection{IndicQARetrieval}
\begin{figure}[hbt!]
    \centering
    \includegraphics[height=8cm,width=7.5cm]{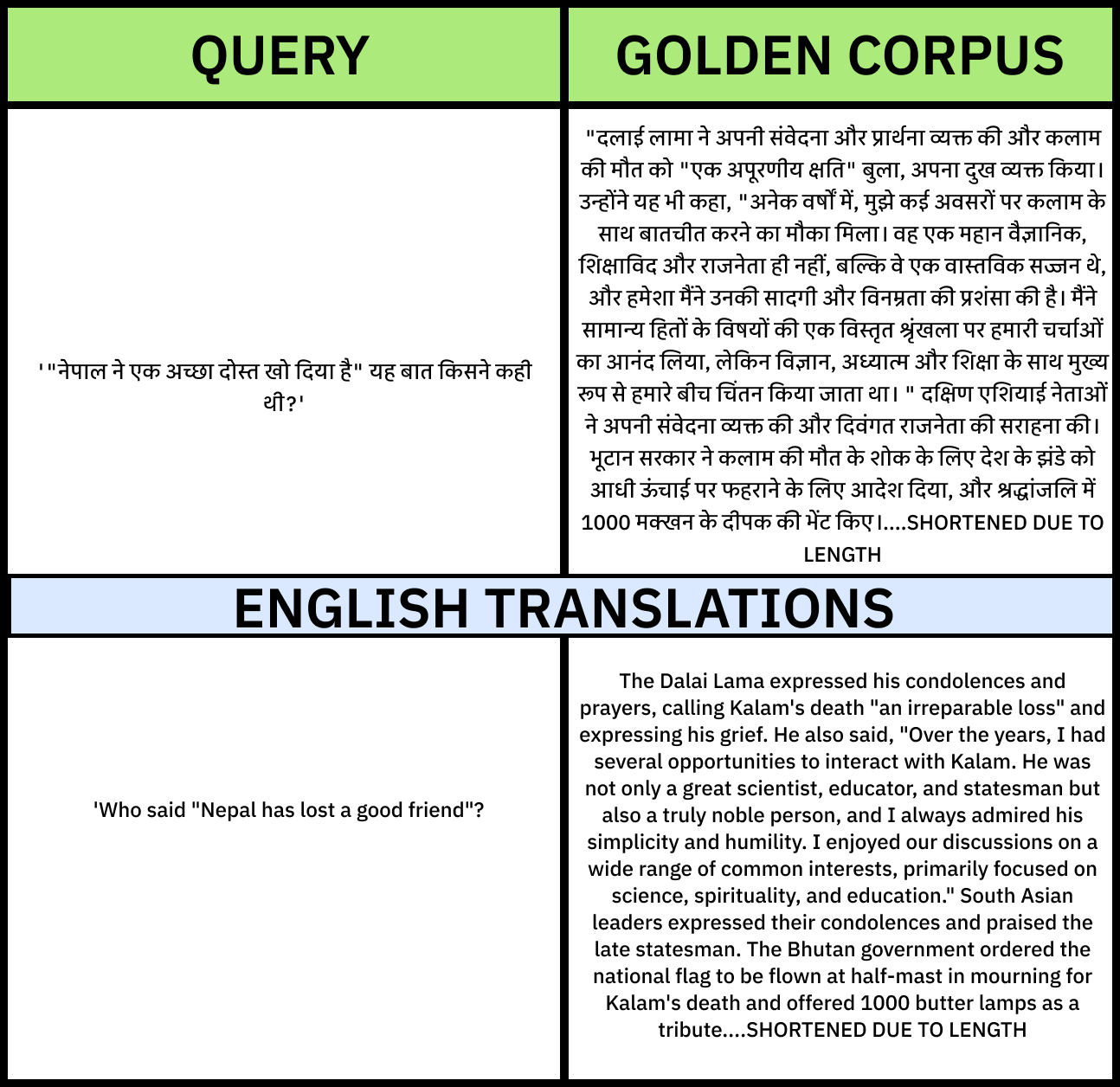}
    \caption{An example of a query with its corresponding golden corpus from the IndicQARetrieval Dataset}
    \label{fig:indicqa_example}
\end{figure}

\begin{enumerate}

\item \textbf{Task Definition: } It is created by transforming the IndicQA dataset \footnote{\url{https://huggingface.co/datasets/ai4bharat/IndicQA}} to a retrieval dataset. The task is similar to NQ, where the model is expected to return the paragraph that contains the answer to the question when given a question. 

    \item \textbf{Domain :} Wikipedia
\end{enumerate}
An example of query with its corresponding golden corpus from the IndicQARetrieval dataset has been provided in Figure \ref{fig:indicqa_example}

\begin{figure}[hbt!]
    \centering
    \includegraphics[height=4cm,width=7.5cm]{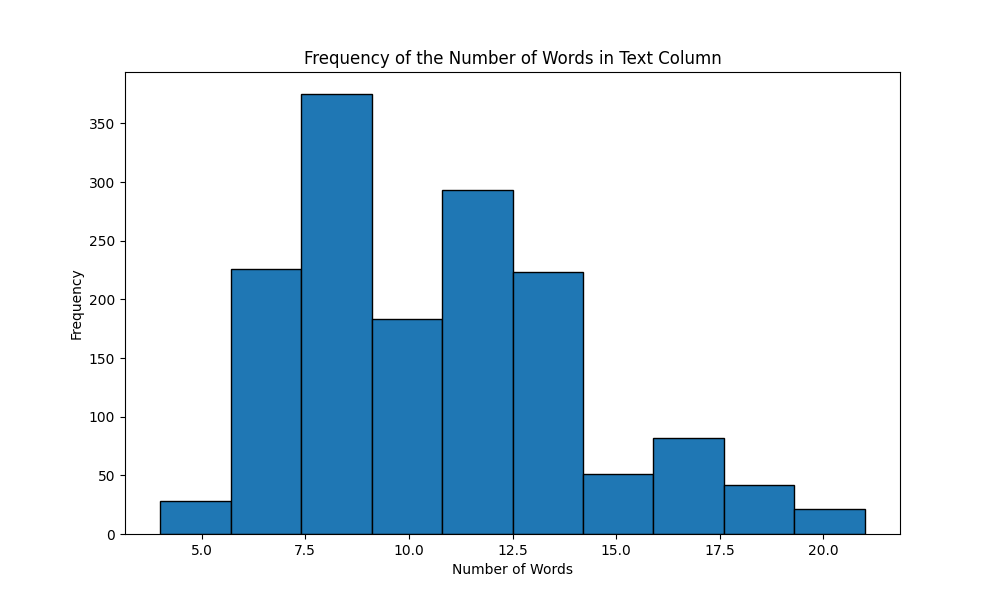}
    \caption{Distribution of the number of words in the queries of IndicQARetrieval Dataset}
    \label{fig:indicqa_queries}
\end{figure}

\begin{figure}[hbt!]
    \centering
    \includegraphics[height=4cm,width=7.5cm]{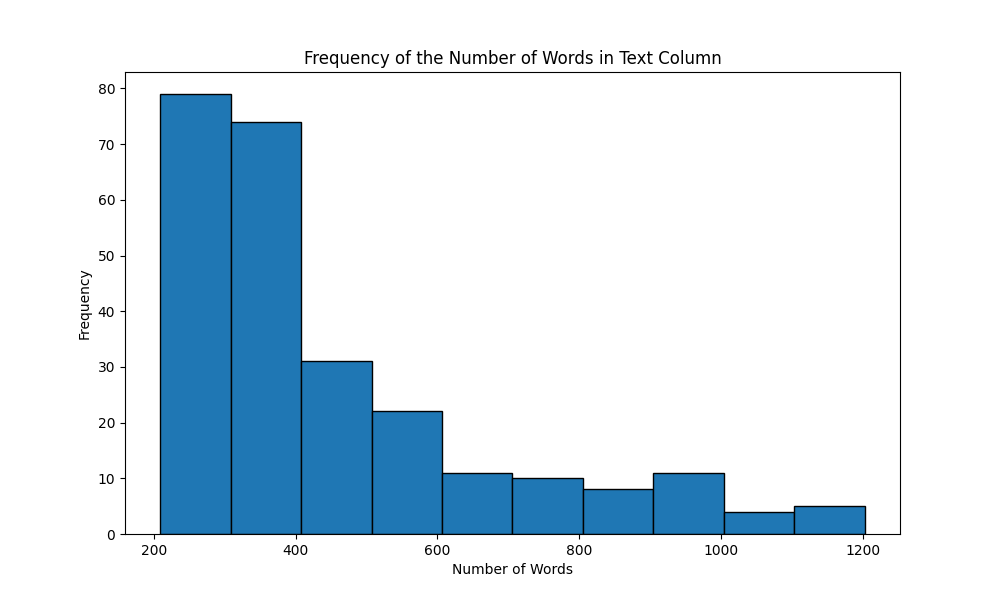}
    \caption{Distribution of Number of Words in the corpus of IndicQARetrieval Dataset}
    \label{fig:indicqa_corpus}
\end{figure}

Distribution of the number of words in the corpus and queries in the IndicQARetrieval dataset has been shown in Figure \ref{fig:indicqa_corpus} and Figure \ref{fig:indicqa_queries} respectively.

\subsubsection{mMARCO}
\begin{figure}[hbt!]
    \centering
    \includegraphics[height=6cm,width=7.5cm]{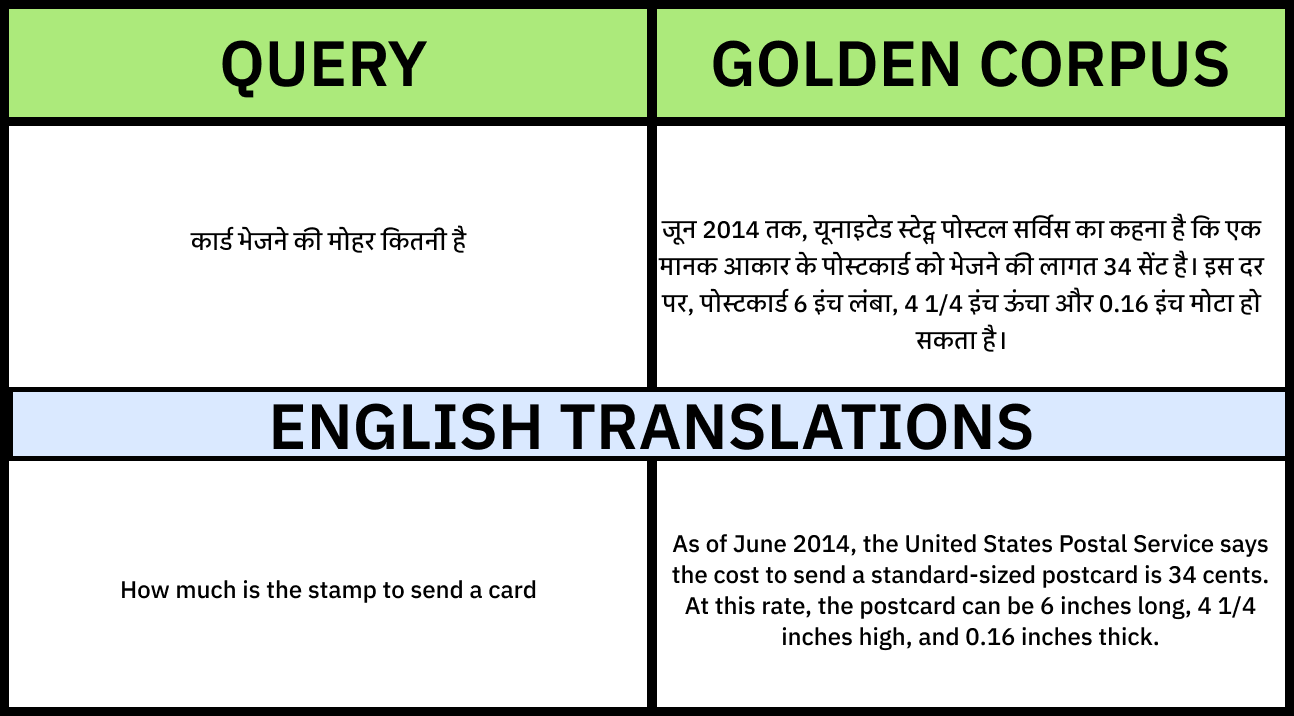}
    \caption{An example of a query with its corresponding golden corpus from the mMARCO Dataset}
    \label{fig:mmarco_example}
\end{figure}

\begin{enumerate}

\item \textbf{Task Definition: } It is a multilingual version of the MSMARCO dataset. The dataset contains the translation of queries from Bing search logs with one text passage from various web sources annotated as relevant.
    \item \textbf{Domain :} Miscellaneous
\end{enumerate}
An example of a query with its corresponding golden corpus from the mMARCO dataset has been provided in Figure \ref{fig:mmarco_example}

\begin{figure}[hbt!]
    \centering
    \includegraphics[height=4cm,width=7.5cm]{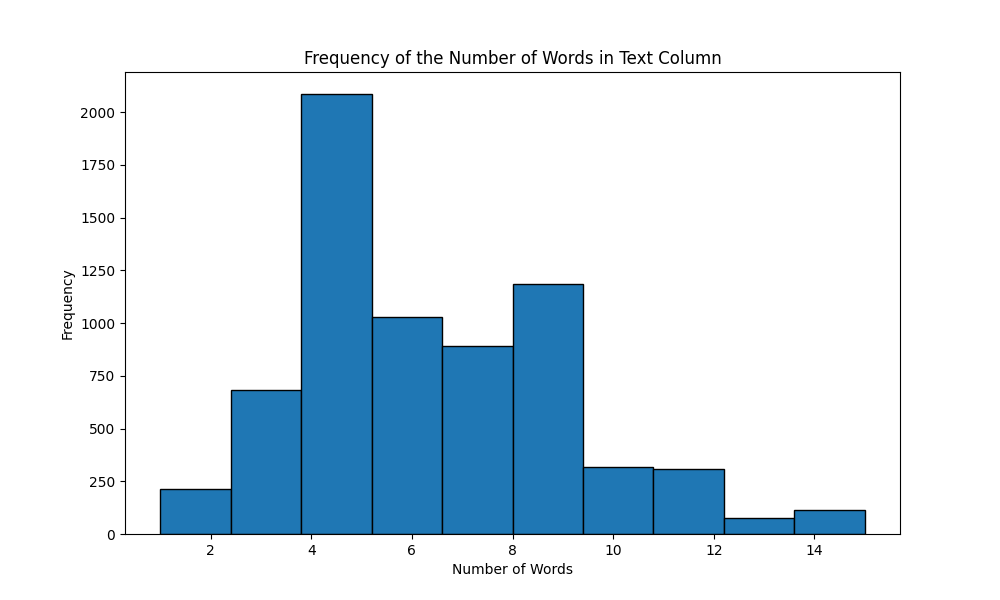}
    \caption{Distribution of the number of words in the queries of mMARCO Dataset}
    \label{fig:mmarco_queries}
\end{figure}

\begin{figure}[hbt!]
    \centering
    \includegraphics[height=4cm,width=7.5cm]{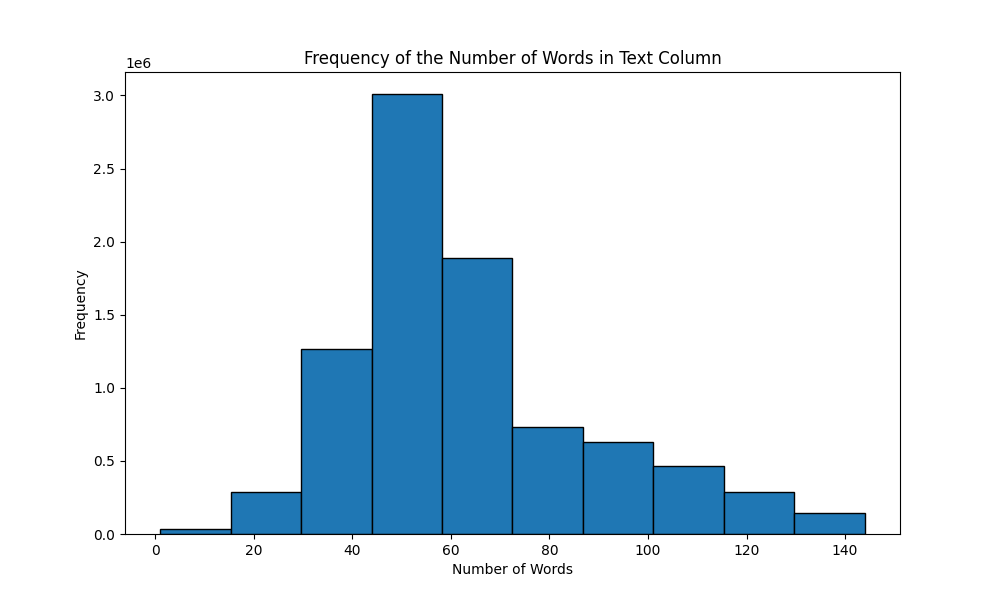}
    \caption{Distribution of Number of Words in the corpus of mMARCO Dataset}
    \label{fig:mmarco_corpus}
\end{figure}

Distribution of the number of words in the corpus and queries in the mMARCO dataset has been shown in Figure \ref{fig:mmarco_corpus} and Figure \ref{fig:mmarco_queries} respectively.

\subsubsection{WikiPediaRetreival}
\begin{figure}[hbt!]
    \centering
    \includegraphics[height=6cm,width=7.5cm]{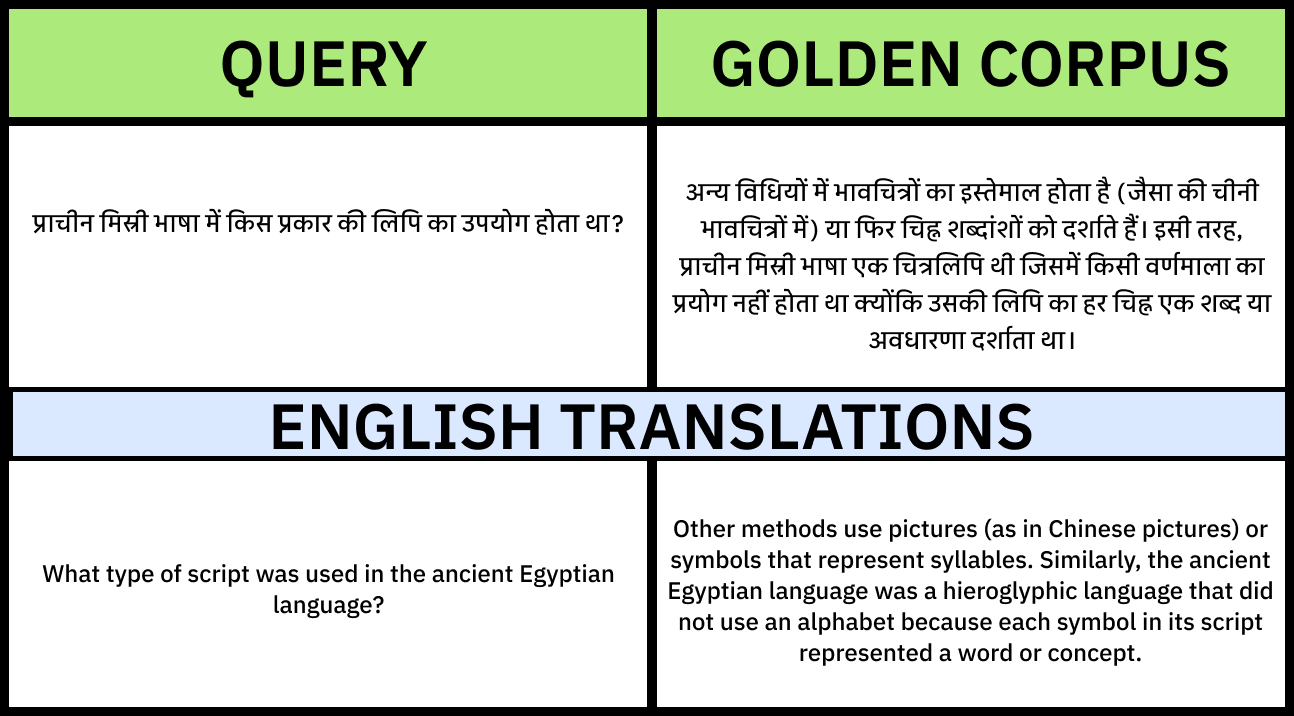}
    \caption{An example of a query with its corresponding golden corpus from the WikiPediaRetreival Dataset}
    \label{fig:wiki_example}
\end{figure}

\begin{enumerate}

\item \textbf{Task Definition: } It is similar to a question-answering task dataset like NQ, where given a query, the model is expected to retrieve a relevant article which answers the question. We have included the Hindi subset of WikiPediaRetrieval dataset \footnote{\url{https://huggingface.co/collections/ellamind/mmteb-6661723dc229e1da8e837cdf}}

    \item \textbf{Domain :} Wikipedia
\end{enumerate}
An example of a query with its corresponding golden corpus from the WikiPediaRetreival dataset has been provided in Figure \ref{fig:wiki_example}

\begin{figure}[hbt!]
    \centering
    \includegraphics[height=4cm,width=7.5cm]{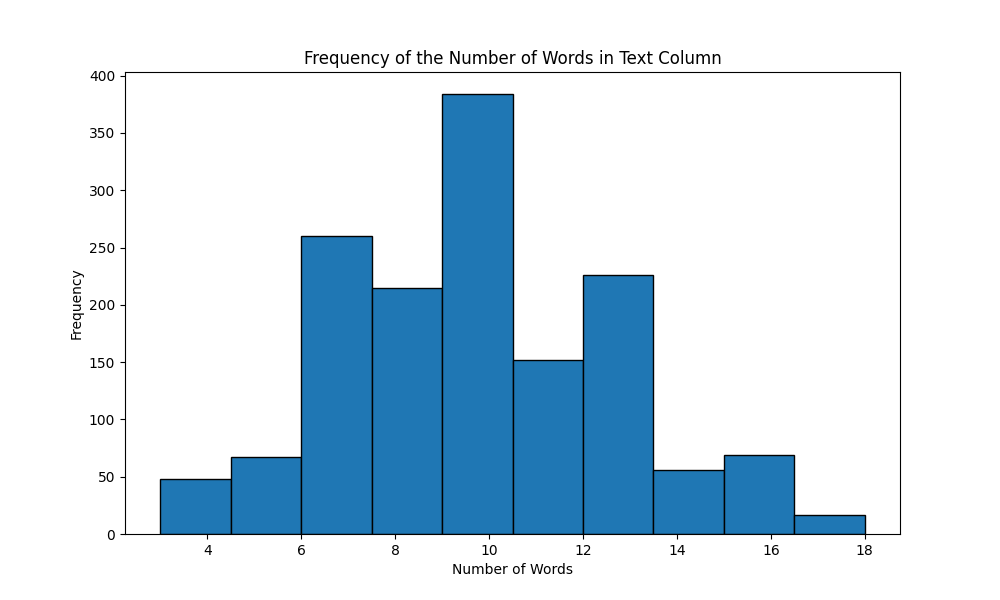}
    \caption{Distribution of the number of words in the queries of WikiPediaRetreival Dataset}
    \label{fig:wiki_queries}
\end{figure}

\begin{figure}[hbt!]
    \centering
    \includegraphics[height=4cm,width=7.5cm]{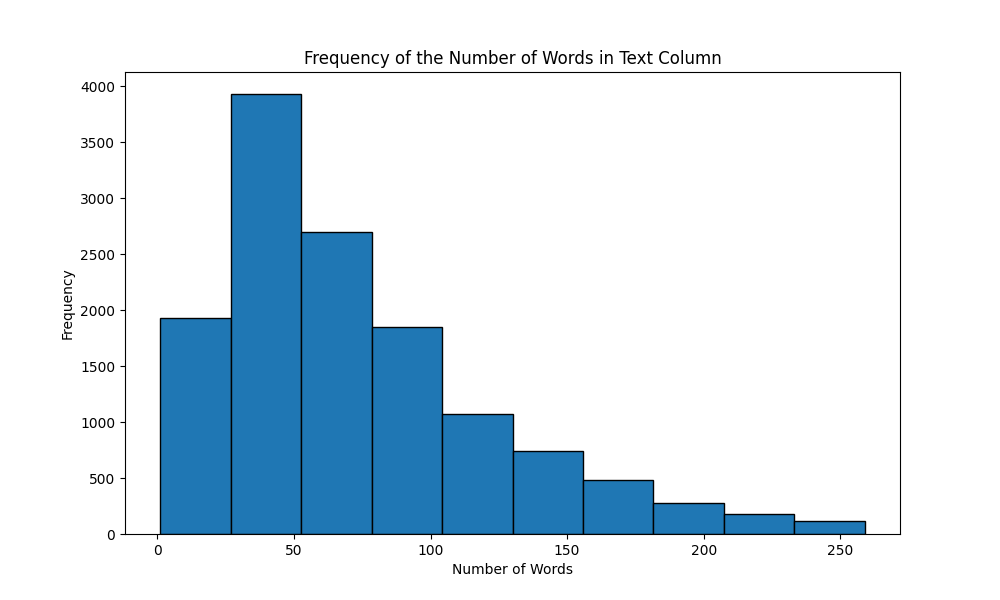}
    \caption{Distribution of Number of Words in the corpus of WikiPediaRetreival Dataset}
    \label{fig:wiki_corpus}
\end{figure}

Distribution of the number of words in the corpus and queries in the WikiPediaRetreival dataset has been shown in Figure \ref{fig:wiki_corpus} and Figure \ref{fig:wiki_queries} respectively.

\subsection{Why choose IndicTrans2 over other available translation models?}
\label{subsec:faq}
\textbf{Ans: }\newcite{gala2023indictrans} clearly illustrate the superior performance of IndicTrans2 over other models and systems like NLLB and Google Translate for  English to Hindi tasks.  In our preliminary analysis on a subset of the BEIR datasets (Chrf scores show in Table \ref{tab:nllbvsindic} ), we also observed that IndicTrans2 outperformed alternative models, such as NLLB, in terms of translation quality.

\begin{table}[!htb]
    \centering
    \begin{tabular}{|c|cc|}
    \hline
        \textbf{Dataset} & \textbf{IndicTrans2} & \textbf{NLLB} \\
    \hline
         Arguana & 56.30 & 37.12 \\
    \hline
         NQ & 76.13 & 60.92 \\
    \hline
    \end{tabular}
    \caption{Chrf scores between the original English text and back translated english text by the respective model for 20,000 randomly selected samples from each dataset.}
    \label{tab:nllbvsindic}
\end{table}

\subsection{Prompts used for Sangraha-IR Query Generation}

The prompt that was used to obtain answers from the Gemini-Flash-1.5 Models is as follows

\textit{Given the above text, generate a question in Hindi which can only be answered by the above passage. The question should be difficult with only semantic similarity and should not contain any lexical overlap; that is, do not directly use the exact phrases as shown in the above passage. You can use synonyms and make the question as complex as possible. Only return the relevant question and nothing else in the format given below :
$<$QUESTION$>$: ...question in hindi.... $</$QUESTION$>$}

\subsection{Ablation of LoRA parameters}
\label{lora}

 \begin{table}[!htb]
    \centering
    \tiny
    \resizebox{0.5\textwidth}{!}{%
    \begin{tabular}{lcc}
    \toprule
      \textbf{Dataset Name}     &
      \multicolumn{1}{c}{\begin{tabular}[c]{@{}c@{}}\textbf{NLLB$_{1.3B (dist)}$-E5}\\ (W/O LoRA, ST)\end{tabular}} & \multicolumn{1}{c}{\begin{tabular}[c]{@{}c@{}}\textbf{NLLB$_{1.3B (dist)}$-E5}\\ (W LoRA, ST)\end{tabular}}
      \\
    \midrule
       ArguAna &  57.24 & 58.19\\
       FiQA-2018  & 34.19 & 34.46\\
       TREC-COVID  & 70.53 & 70.52 \\
       SCIDOCS     & 18.23 & 18.06\\
       SciFact  & 64.36 & 65.23\\
       Touch\'{e}-2020  & 25.21 & 25.74\\
        NQ  & 53.38 & 51.09 \\
        FEVER  & 71.04 & 72.71 \\
        Climate-FEVER  & 22.51 & 23.63 \\
        \midrule
        CC News Retrieval & 31.91 & 29.47\\
        \midrule
        MIRACL & 52.96 & 49.76\\
        IndicQARetrieval  &  62.10 & 61.81\\
        mMARCO & 34.03 & 31.78 \\
       WikiPediaRetrieval &86.20 & 84.36\\
       \midrule
       \textbf{Average} &\textbf{48.84} & 48.34 \\
    \bottomrule
    \hline
    \end{tabular}%
    }
    \caption{NDCG@10 scores of NLLB-E5 model with and without LoRA adapters on E5 model on the Hindi-BEIR Benchmark.}
    \label{tab:lora}
\end{table}

Similar to \newcite{schmidt2024selfdistillationmodelstackingunlocks}, we also experiment with the addition of LoRA adapters during training on top of the E5 model. The LoRA adapters were applied to the "query", "key", and "value" layers of the retrieval model head, with a rank of 32, an alpha parameter of 64, and a dropout rate of 0.05.

Table \ref{tab:lora} presents a side-by-side comparison of the NLLB-E5 model with and without LoRA adapters applied to the E5 head. The results indicate a slight performance boost on certain datasets, particularly those derived from the original BEIR benchmark, which closely resembles the training data. However, it is important to note a significant drop in performance on non-BEIR datasets, such as CC-News Retrieval, MIRACL, IndicQA Retrieval, mMARCO, and Wikipedia Retrieval. We hypothesize that this drop could be due to the LoRA adapters causing the model to overfit on datasets in the style of Sentence Transformers, thereby reducing its generalization capabilities. The overall stronger performance of the model without LoRA adapters is further supported by its higher average NDCG@10 score, and thus, we decided to proceed with the "without LoRA" model setup.

\subsection{Performance of NLLB Encoder}
\label{nllb_baseline}
We trained the NLLB-Encoder using the same knowledge distillation approach as NLLB-E5 and evaluated its performance on subsets of both English and Hindi datasets from BEIR and Hindi-BEIR. The results, presented in Table \ref{tab:nllb_encoder_baseline}, reveal a significant performance gap between NLLB-Encoder and both NLLB-E5 and mE5-Large in the retrieval task. These findings indicate that NLLB-Encoder alone is insufficient to act as a retrieval model and requires additional components, thereby motivating the development of NLLB-E5.

\begin{table}[!htb]
    \centering
    \resizebox{0.5\textwidth}{!}{%
    \begin{tabular}{lccc}
        \toprule
         \textbf{Model}  & \textbf{Hindi ArguAna} & \textbf{Hindi FiQA-2018}  & \textbf{IndicQARetrieval} \\
         \midrule
         NLLB(1.3B)-Encoder & 44.45 & 17.54 & 42.98\\
         NLLB-E5(1.3B) & \textbf{57.24} & \textbf{34.19} & 61.94\\
         mE5-Large & 54.77 & 27.33 & \textbf{67.11}\\
         \bottomrule
    \end{tabular}%
    }
    \caption{NDCG@10 scores of NLLB-Encoder along with mE5-Large and NLLB-E5 on ArguAna and FiQA-2018 from Hindi-BEIR}
    \label{tab:nllb_encoder_baseline}
\end{table}


\end{document}